\documentclass[twocolumn,aps,prb,twocolum,superscriptaddress,bibnotes,amsmath,amssymb,floatfix,footinbib,longbibliography]{revtex4-2}
\usepackage[markup=nocolor, authormarkupposition=left]{changes}
\usepackage{soul}
\usepackage[utf8]{inputenc}
\usepackage[english]{babel}
\usepackage{amsmath,amsfonts,amssymb}
\usepackage[T1]{fontenc}
\usepackage{url}
\usepackage{changes}
\usepackage{amsmath}
\usepackage{amsfonts}
\usepackage{amssymb}
\usepackage{siunitx}
\usepackage{epstopdf}
\usepackage{graphicx}
\graphicspath{{./Figures/}}
\usepackage[
bookmarks=true,
colorlinks=true,
linkcolor=blue,
urlcolor=blue,
citecolor=blue,
pdftex,
bookmarks=true,
linktocpage=true, % makes the page number as hyperlink in table of content
hyperindex=true
]{hyperref}

\newcommand{\SiN}[0]{Si$_3$N$_4$~}

\begin{document}
%\title{Photonic integrated laser with ultra low-noise frequency-agile lasers}
%\title{Photonic integrated lasers with ultra low-noise and frequency agillity}
%\title{Combining ultra low noise and frequency agility in hybrid integrated lasers}
%\title{Ultra low-noise frequency-agile hybrid integrated lasers}
\title{Ultralow-noise frequency-agile photonic integrated lasers}
%\title{Ultra low-noise frequency-agile hybrid integrated lasers for coherent LiDAR}

\author{Grigory Lihachev}
\email{These authors contributed equally}
\affiliation{Institute of Physics, Swiss Federal Institute of Technology Lausanne (EPFL), CH-1015 Lausanne, Switzerland}

\author{Johann Riemensberger}
\email{These authors contributed equally}
\affiliation{Institute of Physics, Swiss Federal Institute of Technology Lausanne (EPFL), CH-1015 Lausanne, Switzerland}

\author{Wenle Weng}
\email{These authors contributed equally}
\affiliation{Institute of Physics, Swiss Federal Institute of Technology Lausanne (EPFL), CH-1015 Lausanne, Switzerland}

\author{Junqiu Liu}
\affiliation{Institute of Physics, Swiss Federal Institute of Technology Lausanne (EPFL), CH-1015 Lausanne, Switzerland}

\author{Hao Tian}
\affiliation{OxideMEMS Lab, Purdue University, 47907 West Lafayette, IN, USA}

\author{Anat Siddharth}
\affiliation{Institute of Physics, Swiss Federal Institute of Technology Lausanne (EPFL), CH-1015 Lausanne, Switzerland}

\author{Viacheslav Snigirev}
\affiliation{Institute of Physics, Swiss Federal Institute of Technology Lausanne (EPFL), CH-1015 Lausanne, Switzerland}

\author{Rui Ning Wang}
\affiliation{Institute of Physics, Swiss Federal Institute of Technology Lausanne (EPFL), CH-1015 Lausanne, Switzerland}

\author{Jijun He}
\affiliation{Institute of Physics, Swiss Federal Institute of Technology Lausanne (EPFL), CH-1015 Lausanne, Switzerland}

\author{Sunil A. Bhave}
\affiliation{OxideMEMS Lab, Purdue University, 47907 West Lafayette, IN, USA}

\author{Tobias J. Kippenberg}
\email[]{tobias.kippenberg@epfl.ch}
\affiliation{Institute of Physics, Swiss Federal Institute of Technology Lausanne (EPFL), CH-1015 Lausanne, Switzerland}

\maketitle
%%%%%%%%%%%%%%%%%%%%%%%%%%%%%%%%%%%%%%%%%%%%%%%%%%%%%%%%%%%%%%%%%%%%%%%%%%%%%%%%
\noindent\textbf{
Low-noise lasers are of central importance in a wide variety of applications, including high-spectral-efficiency coherent communication protocols \cite{armstrong2009ofdm}, distributed fibre sensing \cite{rogers1999distributed}, and long-distance coherent LiDAR \cite{bostick1967carbon}.
In addition to low phase noise, frequency agility, that is, the ability to achieve high-bandwidth actuation of the laser frequency, is imperative for triangular chirping in frequency-modulated continuous-wave (FMCW) based ranging \cite{macdonald1981frequency,uttam1985precision} or any optical phase locking as routinely used in metrology.
While integrated silicon-based lasers have experienced major advances and are now employed on a commercial scale in data centers \cite{kachris2012survey}, %compatible with wafer scale manufacturing,
integrated lasers \textcolor{black}{with sub-100 Hz-level intrinsic linewidth} are based on optical feedback from photonic circuits \cite{fan2020hybrid, jin2021hertz, boller2020hybrid} that lack frequency agility. %, and can only be thermally tuned. %, resulting in the lack of frequency agility.
Here, we demonstrate a wafer-scale-manufacturing-compatible hybrid photonic integrated laser that exhibits ultralow intrinsic linewidth of \textcolor{black}{$25$~Hz} while offering unsurpassed \emph{megahertz} actuation bandwidth, with a tuning range larger than 1~GHz.
Our approach uses ultralow-loss ($1$~dB/m) \SiN photonic microresonators, combined with aluminium nitride (AlN) \textcolor{black}{or lead zirconium titanate (PZT)} microelectromechanical systems (MEMS) based stress-optic actuation. % - both established foundry processes.
Electrically driven low-phase-noise lasing is attained by self-injection locking of an Indium Phosphide (InP) laser chip and only limited by fundamental thermo-refractive noise \textcolor{black}{at mid-range offsets} \cite{huang2019thermorefractive}. %\deleted{allowing both single line operation and microcomb generation}.
By utilizing difference-drive and apodization of the photonic chip, a flat actuation response up to 10~MHz is achieved. % -- orders of magnitude faster piezo-tuning than achieved with bulk lasers.
We leverage this capability to demonstrate a compact coherent LiDAR engine that can generate up to 800~kHz FMCW triangular optical chirp signals, requiring neither any active linearization nor predistortion compensation, and perform a 10~m optical ranging experiment, with a resolution of \textcolor{black}{12.5~cm}.
Our results constitute highly integrated coherent laser system for scenarios where high compactness, fast control, and high spectral purity are required.
It marks an example where hybrid integrated photonics not only achieves a \textcolor{black}{compact} form factor but additionally outperforms traditional bulk external-cavity and fibre lasers in \textcolor{black}{the combination of} frequency agility and phase noise.
%Our work demonstrates that hybrid integrated photonics not only achieves compactness, but moreover outperforms bulk external cavity or fibre lasers in terms of both frequency agility and phase noise.
}

Low-phase-noise lasers are imperative in a wide range of technological and scientific applications, ranging from distributed fibre sensing \cite{rogers1999distributed}, coherent LiDAR \cite{dale2014ultra,hecht2018lidar,rogers2021universal,martin2018photonic,Brandon8693564}, high-spectral-efficiency coherent communication \cite{armstrong2009ofdm,Seimetz4528637} to microwave photonics \cite{marpaung2019integrated}.
In these applications, frequency agility is often a further key requirement, e.\,g., to lock lasers to fibre gratings or achieve cycle-slip-free phase locking.
Over the past decade, the development of heterogeneously integrated lasers has led to a new class of CMOS compatible highly integrated laser sources \cite{duan2014hybrid,komljenovic2016heterogeneous,thomson2016roadmap} that are now commercially employed in data-center interconnects \cite{kachris2012survey}.
%Fundamentally, the linewidth, i.e. the phase noise, of lasers is given by the Schawlow-Townes linewidth limit \cite{schawlow1958infrared,henry1982theory}, which dictates that low-loss laser cavities allow inherently low phase noise.
\textcolor{black}{The fundamental linewidth, i.e., the phase noise, of lasers is given by the modified Schawlow-Townes linewidth limit} \cite{schawlow1958infrared,henry1982theory}\textcolor{black}{, which dictates that low-loss laser cavities with a high number of photons stored in the cavity allow inherently low phase noise.}
In addition to quantum noise, thermodynamical noise, such as thermo-refractive noise due to refractive index fluctuations, constitutes another fundamental limit \cite{braginsky2000thermo,gorodetsky2004fundamental}.
\textcolor{black}{To date, the lowest laser phase noise of compact semiconductor lasers is achieved by self-injection locking with discrete crystalline resonators (sub-Hz white frequency noise level)} \cite{liang2015ultralow,savchenkov2020application}\textcolor{black}{, integrated \SiN microresonators with low confinement waveguides (Hz level)} \cite{jin2021hertz}\textcolor{black}{, gain chips coupled to low-loss PICs with dual Vernier microrings (40 Hz noise level)} \cite{fan2020hybrid}\textcolor{black}{, single microring (13 kHz noise level) }\cite{Stern:17}\textcolor{black}{~or other configurations (sub-kHz and kHz noise level) }\cite{huang2019high,tran2019tutorial,Lin8374920}\textcolor{black}{, that have very limited frequency agility.} The main difficulty in achieving high frequency agility lies in rapid frequency tuning of the photonic resonator that provides laser linewidth narrowing while maintaining a long photon lifetime in the resonator.

Here we overcome this outstanding challenge.
By using self-injection locking (SIL) of an III-V InP laser to an ultralow-loss CMOS-compatible Si$_3$N$_4$ microresonator \cite{jin2021hertz,puckett2021422,liu2020high} monolithically integrated with AlN MEMS-based actuators \cite{tian2020hybrid,liu2020monolithic}, we achieve both frequency agility and ultra-narrow linewidth, exhibiting phase noise that is on par with fibre lasers - the workhorse for fibre sensing.
% with  to apply self-injection locking technique to an InP laser chip.
Owing to the high $Q$ of the Si$_3$N$_4$ microresonator resonances, the InP laser shows an intrinsic linewidth of \textcolor{black}{$\sim25$~Hz}.
Using the AlN piezoelectrical actuators engineered based on novel contour mode cancellation and differential drive schemes allows the photonic microresonator to be frequency-modulated via the stress-optic effect with a flattened response up to the actuation frequency of 10~MHz \textcolor{black}{- order of magnitude improvement} \cite{liang2018low} \textcolor{black}{due to the planar co-integration}.
This enables a class of compact LiDAR sources that do not require external linearization of the FMCW signal.
\textcolor{black}{We generate narrow-linewidth triangularly chirped lasers, at rates of up to 800 kHz and nonlinearities as low as 1\% without digital predistortion or complex direct microwave signal synthesis and perform an FMCW LiDAR demonstration at 100 kHz chirp frequency.}
The versatility permitted by the optical and mechanical properties of the system shows great promise in applications including high-bit-rate telecommunication \cite{marin2017microresonator}, field-deployable frequency referencing \cite{truong2016accurate}, frequency-agile rapid-scanning spectroscopy \cite{millot2016frequency,debecker2005high}, and low-cost FMCW LiDAR engines \cite{liu2020monolithic}.

\section*{Results}

%\subsection*{Hybrid self-injection-locked laser system}
%%%%%%%%%%%%%%%%%%%%%%%%%%%%%%%%%%%
\begin{figure*}[t!]
\centering
\includegraphics[width=2\columnwidth]{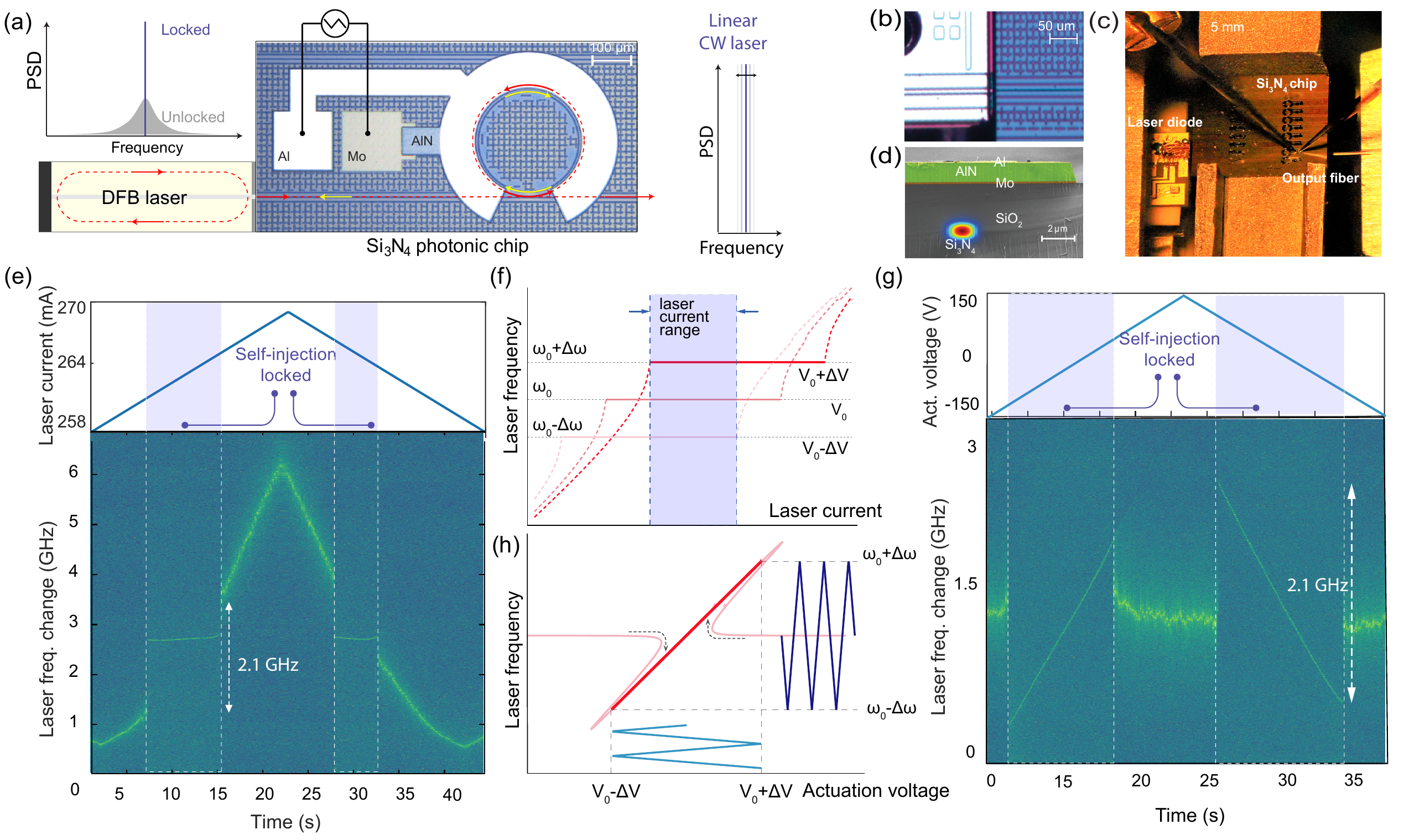}
\caption{
\textbf{Schematic of the hybrid integrated laser system.}
(a)~Principle of laser linewidth narrowing via laser self-injection locking.
The laser frequency tuning is realized by applying a sweeping electrical signal on the monolithically integrated AlN actuator.
(b)~Optical micrograph showing the DFB laser butt-coupled to the Si$_3$N$_4$ photonic chip.
(c)~Photo of the experimental setup with DFB laser (left), Si$_3$N$_4$ chip (middle), output lensed fibre, probes for piezoactuator (top).
(d)~False-colored scanning electron microscope (SEM) image of the sample cross-section, showing the piezoelectric actuator integrated on the Si$_3$N$_4$ photonic circuit.
The piezoelectric actuator is composed of Al (yellow), AlN (green) and Mo (red) layers on top of Si$_3$N$_4$ buried in SiO$_2$ cladding.
(e)~Spectrogram showing laser frequency change upon the linear tuning of the diode current, \textcolor{black}{measured for 190.7 GHz FSR microresonator}, dashed areas correspond to the range where the laser is self-injection locked (featured with minimal lasing frequency fluctuations).
(f) Schematic of the tuning of the laser frequency.
Different voltage levels applied to the piezoactuator correspond to different microresonator resonance frequencies, thus leading to the different frequencies of the laser when the laser current is in the range for self-injection locking.
(g) Spectrogram of laser frequency change upon the linear tuning of the cavity resonance by piezoelectric actuator, \textcolor{black}{measured for 190.7 GHz FSR microresonator},  dashed areas correspond to the range where the laser is self-injection locked to the shifting resonance.
(h) Schematic of linear laser frequency tuning with integrated piezoactuator.
By applying the triangular voltage ramp to the piezoactuator, we transduce the cavity resonance shift induced by the piezoactuator to the triangular laser frequency change while operating inside the locking range.
}
\label{Fig:Fig1}
\end{figure*}
%%%%%%%%%%%%%%%%%%%%%%%%%%%%%%%%%%%

\noindent \textbf{Hybrid self-injection-locked laser system}.
As illustrated in Fig. \ref{Fig:Fig1}(a), the hybrid laser system comprises a III-V laser chip with a distributed feedback (DFB) structure and a photonic chip-based ultralow-loss Si$_3$N$_4$ resonator with a monolithically integrated AlN piezoelectrical actuator.
The Si$_3$N$_4$ photonic chips are fabricated using the photonic Damascene reflow process \cite{pfeiffer2016photonic,liu2018ultralow,liu2020high}, and feature \textcolor{black}{intrinsic quality factor} $Q_0>10\times10^6$.
\textcolor{black}{Frequency-dependent transmission, reflection and cavity linewidth data is presented in the SI Fig. 5 for all chips used in this work}.
Made from polycrystalline AlN as the main piezoelectric material, the actuator has molybdenum (Mo) and aluminium (Al) as the bottom
(ground) and the top electrodes, respectively \cite{tian2020hybrid} as shown in Fig.~\ref{Fig:Fig1}(d).
Applying a voltage between the electrodes tunes the microresonator frequency via the stress-optic effect \cite{huang2003stress}.
The DFB laser diode is butt-coupled to the Si$_3$N$_4$ photonic chip as shown in Fig.~\ref{Fig:Fig1}(b,c) operating at a lasing wavelength of 1556~nm with \textcolor{black}{a free space} output power of 80~mW. \textcolor{black}{Such hybrid packaging approach} \cite{shen2020integrated} \textcolor{black}{and also heterogeneous integration with InP laser and \SiN PIC fabricated on a single silicon substrate} \cite{xiang2021laser}\textcolor{black}{, have recently been demonstrated with \SiN microresonators.}
By tuning the current of the laser diode, we sweep the relative frequency between the laser and the resonator modes to attain self-injection locking via the coupling of counter-propagating microresonator modes induced by Rayleigh-backscattering \cite{kippenberg2002modal} (cf. SI and Fig.~\ref{Fig:Fig1}(e)).
The gap between the laser chip and the Si$_3$N$_4$ photonic chip is adjusted for optimal feedback phase \cite{kondratiev2017self}, which yields the maximum self-injection locking range of up to 2.1~GHz (cf. Fig.~\ref{Fig:Fig1}(e)).
Self-injection locking in this manner has been attained in previous work, however, the ultralow-loss photonic integrated resonator enables the substantial reduction in the phase noise via self-injection locking that is hitherto only surpassed in crystalline microresonators \cite{liang2015ultralow}. \textcolor{black}{In our work we only consider linear regime of laser operation, however soliton microcomb formation is also possible in our devices (see SI)}.
In order to achieve frequency agility, we bias the diode in the center of the locking plateau.
In this manner, changes in the microresonator frequency will maintain injection locked operation and therefore lead to frequency tuning (cf. Fig.~\ref{Fig:Fig1}(f)) while maintaining injection locking.
The AlN actuator will therefore transduce the applied voltage directly to changes in the optical frequency.
Figure~\ref{Fig:Fig1}(h) shows the range over which the photonic resonator frequency can be tuned while still maintaining injection locking (corresponding to a tuning range of up to 2.1~GHz).
%By either tuning the current of the laser diode or the voltage applied to the AlN actuator, we sweep the relative frequency between the laser and the resonator modes to attain the state of self-injection locking with the Rayleigh-backscattered light from the microresonator. Dependent on the on-chip laser power and the parametric oscillation threshold in the microresonator, single-line lasing or microcomb generation (see SI) can be obtained from the output of the Si$_3$N$_4$ photonic chip. While the self-injection-locked state is maintained, one can apply triangular voltage ramps to the AlN actuator to execute agile frequency tuning of the laser.

%%%%%%%%%%%%%%%%%%%%%%%%%%%%%%%%%%%%%%%%%%%%%

\begin{figure*}[t!]
\centering
\includegraphics[width=2\columnwidth]{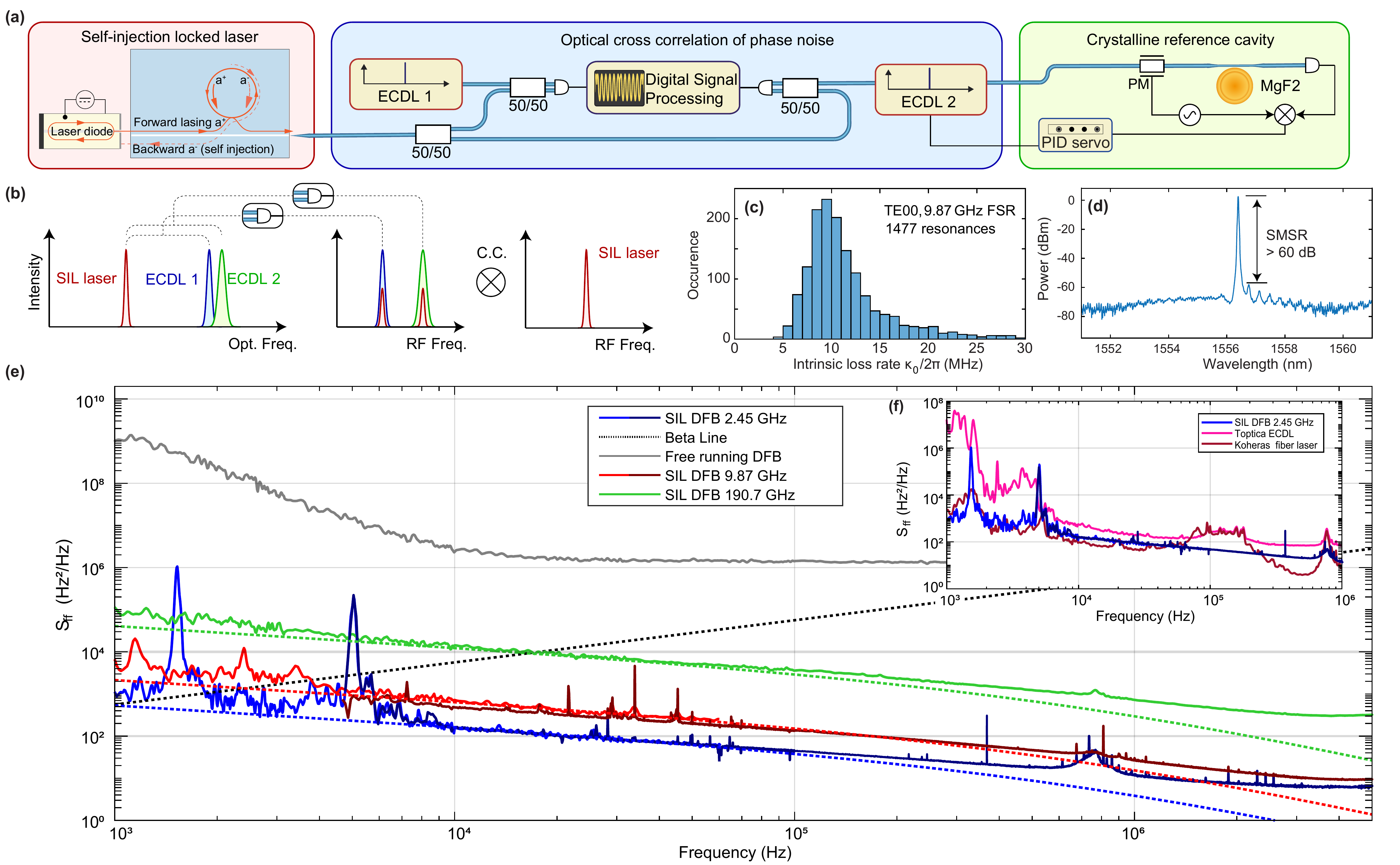}
\caption{
\textbf{Spectral purity of the self-injection-locked laser.}
(a) Experimental schemes of laser frequency noise measurements, using optical cross-correlation technique, or heterodyne beat with the reference laser (ECDL PDH-locked to a high-$Q$ crystalline resonator).
(b) Schematic in the frequency domain of optical cross-correlation (C.C.) technique.
(c) The Si$_3$N$_4$ photonic chips with an FSR of 9.87 GHz have an intrinsic loss of $\kappa_0/2\pi<10$ MHz, corresponding to a quality factor of $Q_0>20\times10^6$.
The $\kappa_0/2\pi$ histogram of 1477~TE$_{00}$ resonances from a 9.87 GHz FSR microresonator is shown.
(d) Optical spectrum of the SIL DFB emission.
(e) \textcolor{black}{Single-sided PSD of frequency noise} of the hybrid integrated laser system upon self-injection locking to microresonators with FSRs: 190.7~GHz (green), 9.87~GHz (red), 2.45~GHz (blue) and free running regime (grey).
Dark color traces correspond to optical cross-correlation data, light color traces to the heterodyne beat method.
The dotted red, green, blue lines indicate the calculated thermo-refractive noise limit for \SiN microresonators with different FSR.
(f) The inset shows a comparison of laser frequency noise of SIL DFB and ECDL (Toptica CTL), fibre laser (NKT Koheras \textcolor{black}{Adjustik E15}).
}
\label{Fig:Fig2}
\end{figure*}
%%%%%%%%%%%%%%%%%%%%%%%%%%%%%%%%%%%%%%%%%%%%%%%%%%%%%%%%%%%%%%%%%%%%%%%%%%%%%%%%

\noindent \textbf{Ultralow optical phase noise}.
Since the laser linewidth narrowing factor, i.e. the ratio of the free-running laser linewidth to the linewidth of the injection-locked laser, is quadratically proportional to the $Q$ of the resonator mode to which the laser is self-injection-locked \cite{kondratiev2017self}, the high loaded $Q$ of the Si$_3$N$_4$ microresonator \textcolor{black}{(see Fig.~{\ref{Fig:Fig2}}(c))} can significantly reduce the optical linewidth and improve the side mode suppression ratio (SMSR).
We present simulations of the self-injection locking dynamics in the SI.
Figure \ref{Fig:Fig2}(d) shows the optical spectrum of the laser while it is self-injection locked, demonstrating an SMSR of 60~dB.
The relative intensity noise (RIN) of the laser is displayed in the SI.
%The laser power through the bus waveguide in the Si$_3$N$_4$ photonic chip is mixed with an ultra-narrow-linewidth laser (whose phase noise is much lower than that of the self-injection-locked laser under test here) and registered by a low-noise photodetector. By measuring the frequency (phase) noise of the beat signal we can infer the frequency noise of the hybrid integrated laser (see Methods for the details of the measurement).
To measure and confirm the ultralow phase noise of the laser, we adopt two approaches \textcolor{black}{(see Figure {\ref{Fig:Fig2}}(a))} to measure the frequency noise power spectral density $S_{\nu}(\Omega)$ (single-sided PSD, in units of $\mathrm{Hz^2/Hz}$), including 1) beating the injection-locked laser with an narrow-linewidth reference laser \textcolor{black}{(see the SI for the reference laser details)} and measuring the beat signal's frequency noise spectrum with an electric spectrum analyser (ESA), and 2) optical cross-correlation-based noise spectrum characterization \cite{xie2017phase} using two auxiliary lasers.
(cf. Methods for details \textcolor{black}{and Figure {\ref{Fig:Fig2}(b)}}).
Figure \ref{Fig:Fig2}(e) shows the frequency noise spectra of the hybrid lasers in free-running condition and self-injection locking states, respectively.
For the injection locking, three microresonators with distinct sizes and different free spectral ranges (FSRs) of 190.7, 9.87 and 2.45~GHz are tested.
In general, the self-injection locking suppresses the frequency noise of the laser by more than 30~dB across the entire spectrum.
At frequencies below 1~kHz, \textcolor{black}{technical noises due the ambient temperature fluctuations, and the coupling gap instabilities caused by acoustic vibrations (which can be eliminated by packaging) leads to hybrid integrated laser RIN and also to a transduction to frequency noise at offsets < 1 kHz (see SI).}
%At frequencies above 1 MHz, the laser frequency noise reaches a plateau of only 50~Hz$^2/$Hz for the 2-GHz \SiN device.
%The optical cross-correlation-based characterization with its ultra-low measurement noise floor allows us to confirm that the laser intrinsic linewidth is below 40~Hz (i.e., a white noise floor of $\sim8$~Hz$^2/$Hz).
%The measurement is repeated for three different resonators with an FSR of 190, 10 and 2~GHz and the frequency noise spectral density found to be in excellent quantitative agreement with thermo-refractive noise (TRN) limit \cite{huang2019thermorefractive} of the Si$_3$N$_4$ microresonator.
\textcolor{black}{The optical crosscorrelation-based characterization reveals that the laser frequency noise reaches a plateau (white noise floor) of only 8~Hz$^2/$Hz for the 2.45 GHz device, 10~Hz$^2/$Hz for the 9.87 GHz device at 3 MHz offset.} \textcolor{black}{The frequency noise power spectral density was found to be in good quantitative agreement with fundamental thermo-refractive noise (TRN) limit} \cite{huang2019thermorefractive} \textcolor{black}{of the \SiN microresonator at mid range offsets from 5 to 100 kHz.}
\textcolor{black}{Thus, we demonstrate using high confinement \SiN platform the laser performance limited by thermo-refractive noise, which has only been shown in \SiN low-confinment waveguides} \cite{jin2021hertz}.
\textcolor{black}{We note that this intrinsic linewidth can be further reduced by using microresonators with even larger optical mode volume and therefore reduced TRN.}    %(thus larger effective mode volume) since theory \cite{kondratiev2017self} suggests that the measured $Q$ factor of the microresonator permits even larger linewidth narrowing factor.
To illustrate the performance of this integrated laser, we compare the laser frequency noise to two commonly used lasers.
Fig.\ref{Fig:Fig2} (f) reveals that our hybrid integrated laser is better than a state of the art commercial ECDL (Toptica CTL) and on par with a commercial fibre laser (NKT Koheras \textcolor{black}{Adjustik E15}) at the offset frequency range of 1--50 kHz.
Another method to quantify and compare the linewidth from the measured frequency noise spectral density $S_{\nu}(f)$ is to invoke the beta-line \cite{DiDomenico:10}.
By integrating frequency noise up to the frequency of the interception point with a line $S_{\nu}(f)=8\cdot$ln(2)$f/\pi^2$ we obtain an area A which we use for FWHM measure of the linewidth (8$\cdot$ln(2)A)$^{1/2}$, and from Fig. \ref{Fig:Fig2} (e) we infer this value to be equal to 46 kHz for the 9.87 GHz FSR \SiN device. %and 238 kHz for 2 GHz device.
%\textcolor{black}{Lets call all noise spectral densities with one subscript only, since they are classical.}
%%%%%%%%%%%%%%%%%%%%%%%%%%%%%%%%%%%%%%%%%%%%%

%%%%%%%%%%%%%%%%%%%%%%%%%%%%%%%%%%%
\begin{figure*}[t!]
\centering
\includegraphics[width=\textwidth]{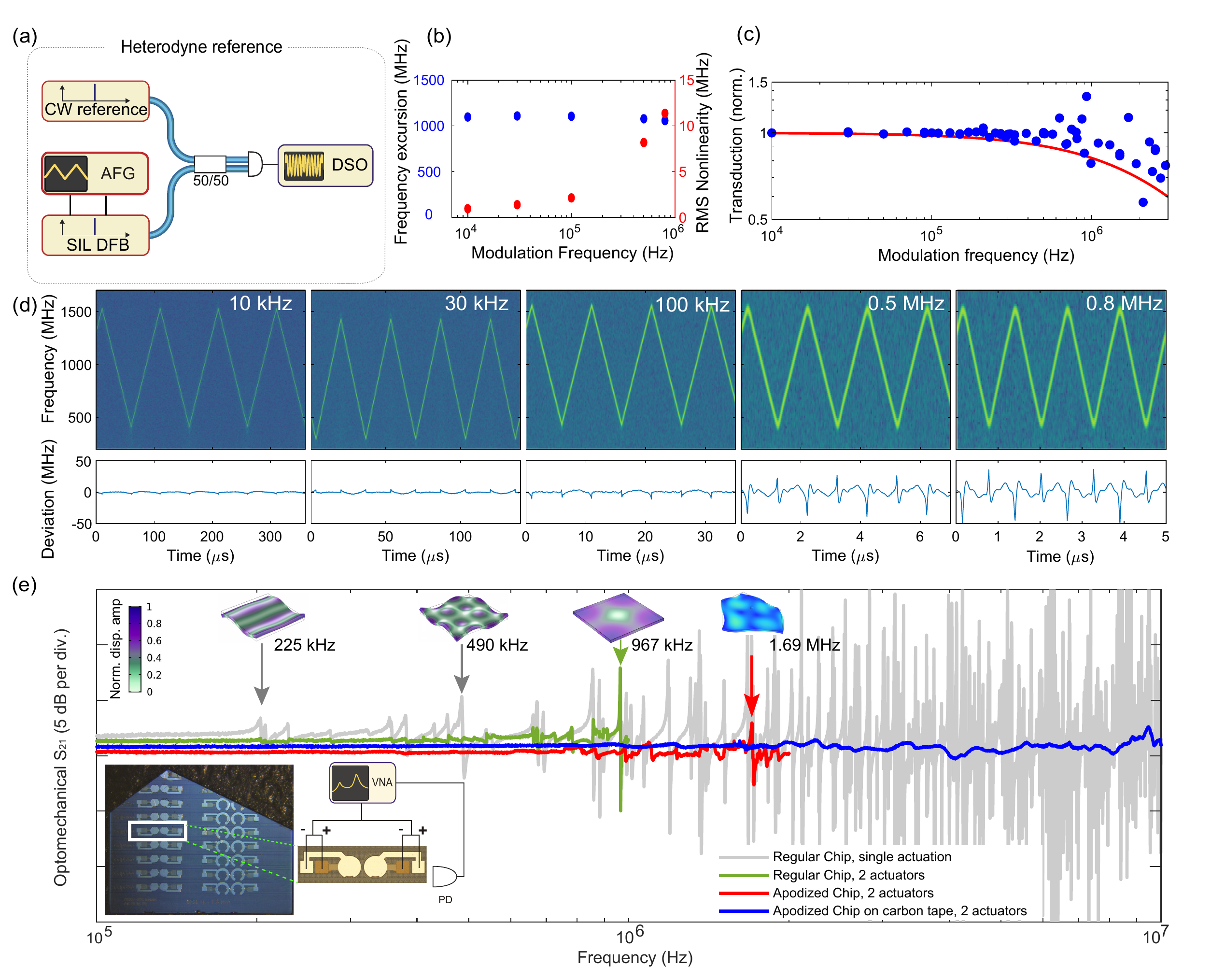}
\caption{
\textbf{Frequency-agile tuning with integrated AlN piezoactuator.}
(a) Experimental setup for heterodyne beat note characterization of the frequency-agile hybrid-integrated laser.
A continuous-wave (CW) external-cavity diode laser is used as a reference, and the beatnote is recorded on a fast oscilloscope (DSO) and analyzed with short-time Fourier transforms.
(b) Frequency excursion (blue) and residual root-mean-square (RMS) nonlinearity (red) of triangular laser chirps.
The AlN actuator is driven with a peak-to-peak amplitude of 150~V.
(c) Piezo-voltage to laser frequency transduction is calculated from the harmonic spectral content of the laser frequency chirp and the best fitted perfect triangular frequency chirp (red).
(d) Time-frequency spectrogram of the heterodyne beat-notes for different triangular chirp repetition frequencies.
Bottom row: Residual of least-squares fitting of the time-frequency traces with symmetric triangular chirp pattern.
(e) Suppression of photonic chip mechanical resonances.
Measured responses of the stress-optic actuation for 190.7~GHz FSR microresonator using disk-shaped piezoactuator with single actuation (grey), dual actuators with difference-actuation for a square Si$_3$N$_4$ chip (green), an apodized chip (red) and an apodized chip on a carbon tape (blue).
Inset: \textcolor{black}{three mechanical modes of regular Si$_3$N$_4$ chip (225 kHz, 490 kHz, 967 kHz) and apodized  chip (1.69 MHz) simulated with FEM}, eigenfrequencies denoted by the arrows with displacement amplitude profile visualization.
Photo of the apodized chip with the dual-actuator configuration, \textcolor{black}{exact chip dimensions are provided in the SI.}
Experimental schematic for difference driving of dual actuator.
}
\label{Fig:Fig3}
\end{figure*}
%%%%%%%%%%%%%%%%%%%%%%%%%%%%%%%%%%%%%%%%%%%%%%%%%%%%%%%%%%%%%%%%%%%%%%%%%%%%%%%%
\noindent \textbf{Frequency-agile tuning}.
We next demonstrate the frequency agility of our hybrid laser.
To this end, we carry out frequency-modulation of the self-injection-locked laser by applying a time-varying voltage to the integrated AlN actuator (cf. Fig. \ref{Fig:Fig3} (a)).
Triangular ramp signals of ramping frequencies from 10~kHz to 800~kHz are generated with an arbitrary frequency generator and amplified to 150~$\mathrm{V}_\mathrm{pp}$.
In the self-injection locking range, the change of microresonator frequency imprints directly on the laser output frequency, even without additional feedback on the pump current of the laser.
The time-varying laser output frequency is characterized by measuring a heterodyne beatnote with a reference ECDL \textcolor{black}{(free running Toptica CTL)} on a fast photodetector. \textcolor{black}{We define chirp nonlinearity as the root mean square (RMS) deviation of the measured frequency tuning curve from a perfect triangular ramp that is determined with least-squares fitting. The phase noise PSD of the tuned laser can also be directly retrieved by Hilbert's transform from the heterodyne beat note (cf. SI)}.
Figure~\ref{Fig:Fig3} (b) and (c) summarize the main results of the heterodyne beat experiment with the SIL laser locked to a 190.7~GHz microresonator. %The 3-dB bandwidth of the triangular ramp modulation is higher than 20~MHz.
%Compared to the AlN disc actuator used in earlier work \cite{liu2020monolithic}, the tuning efficiency of the donut actuator is reduced by $\sim50\%$. Yet,
The large tuning range of $>1$~GHz at high ramping speeds up to \textcolor{black}{800~kHz}, with small chirping RMS nonlinearities \textcolor{black}{down to} 1~$\%$ as shown in Fig.~\ref{Fig:Fig3} (b) showcases the remarkable frequency agility of our system.
This excellent linearity and the almost vanishing hysteresis of the monolithically integrated AlN actuator \cite{liu2020monolithic} facilitates the generation of highly linear triangular chirps for modulation frequencies up to 100~kHz without the need for active or passive linearization.
Figure~\ref{Fig:Fig3}(d) presents the processed laser frequency spectrograms and the corresponding nonlinearities at five different ramping frequencies respectively.
At 10~kHz modulation frequency, the achieved RMS nonlinearity is as low as 600~kHz (relative nonlinearity $5\cdot 10^{-4}$), which only degrades slightly for 100~kHz tuning rate to 1.5~MHz.
\textcolor{black}{In Figure~{\ref{Fig:Fig3}}(c), we plot the frequency-dependent transduction from the frequency modulation amplitude of the first 17 harmonics of each modulation frequency (10 kHz - 800 kHz) from the experimental data presented in Fig.~{\ref{Fig:Fig3}}(d). The peak around 900 kHz matches well with the first mechanical mode of the chip presented in Fig.~{\ref{Fig:Fig3}}(e). The 4 MHz low pass cut-off of the high voltage amplifier is indicated in red.}
Together with more than 1~GHz tuning range, such a high actuation bandwidth exceeds the performance of common benchtop laser systems that rely on bulk piezo or electrooptic components, making this laser source an ideal candidate for direct implementation in long range FMCW LiDAR systems \cite{feneyrou2018novel}, that can operate at rates reaching megapixel-per-second.

\noindent \textbf{Flattening actuation response via photonic chip mechanical modes suppression}.
To achieve the optimal performance of the hybrid integrated laser as an FMCW LiDAR engine at high measurement rates, in addition to a large actuation bandwidth B (which determines LiDAR resolution $c/2B$), a flattened actuation transfer function is highly desired for minimizing chirping nonlinearity.
The inset of Figure~\ref{Fig:Fig3}(e) \textcolor{black}{shows the setup of the actuation response measurement.
In this measurement, the actuation voltage derived from a vector network analyser (VNA) is applied on the actuator, and a laser is frequency-tuned to sit on the side of resonance, to measure the response.} %The frequency modulation due to the actuation is converted to the intensity modulation of the transmitted laser light that is received by a fast photodetector.
Figure~\ref{Fig:Fig3}(e) presents the measured optomechanical response of the single-actuator configuration (grey) and the dual-actuator configuration (green, red, blue).
As shown in Fig. \ref{Fig:Fig3}(e) (grey), the fabricated \textcolor{black}{AlN} piezoelectrical monolithic actuator \textcolor{black}{excites many mechanical bulk or contour modes of the photonic chip}, leading to a nonflat actuation response.
The inset of Fig.~\ref{Fig:Fig3} (e) shows finite element simulations of the flexural modes of the photonic chip, that match the observed actuation resonances.
The increasing mode density of the Si$_3$N$_4$ photonic chip with actuation frequency severely limits the flat effective actuation bandwidth.
We mitigate this effect, first, by developing the active cancellation scheme with a difference-actuation.
In this scheme, an additional AlN actuator with the same geometry is fabricated adjacent to the microresonator.
The two actuators are driven by the same signal but with a 180-degree phase shift to cancel the actuation of the mechanical modes of the photonic chip.
As a result, while the stress-optic effect exerted on the microresonator is the same, the excitation of mechanical resonances can be effectively suppressed as shown in Fig.~\ref{Fig:Fig3}(e) (green).
The scheme effectively reduces modes below 1~MHz, mainly cancelling the flexural modes due to far-field destructive interference, a scheme inspired by nanomechanical membranes \cite{wilson2011high}.
Many mechanical modes of relatively low resonance frequencies are flexural modes whose vibrations are caused by transverse standing waves.
The bulk mechanical modes whose vibrations are caused by longitudinal standing waves (displacement amplitude profile visualization shown at 967~kHz frequency) can be eliminated by judiciously shaping the geometry of the photonic chips \cite{burak2017acoustic}. To improve the actuation further, and suppress the bulk acoustic modes, we next apodized the photonic Si$_3$N$_4$ chip \textcolor{black}{by dicing the released chip}. We observe a reduction in the number of bulk mechanical modes in an apodized photonic chip.
Figure~\ref{Fig:Fig3} (e) (red) shows that the mechanical resonances below 1~MHz are completely removed with \textcolor{black}{the first resonance of apodized chip at 1.69 MHz, matching the FEM simulation.}
We further flatten the actuation response by attaching the apodized chip on a piece of carbon tape and then differentially driving the actuators as explained before.
In this way, both the flexural and the bulk mechanical modes are damped up to the first \textcolor{black}{fundamental high-overtone bulk acoustic resonator mode (HBAR)} at 17~MHz \cite{liu2020monolithic}. % It is also observed that even without using the difference-actuation scheme all the mechanical resonances of the apodized chip are significantly suppressed till the first HBAR mode
The active cancellation scheme on an apodized chip placed over a carbon tape limits the variations of the actuation response within 1 dB yielding a record flat response bandwidth of nearly 20~MHz \textcolor{black}{(see the SI for the full plot)}.
Such flat actuation response can improve the linear chirping performance of the FMCW LiDAR, and in particular allows an increase of the FMCW signal frequency, thereby directly increase the speed of the LiDAR to beyond megapixel-per-second rates.

\begin{figure*}[hbt]
\centering
\includegraphics[width=2\columnwidth]{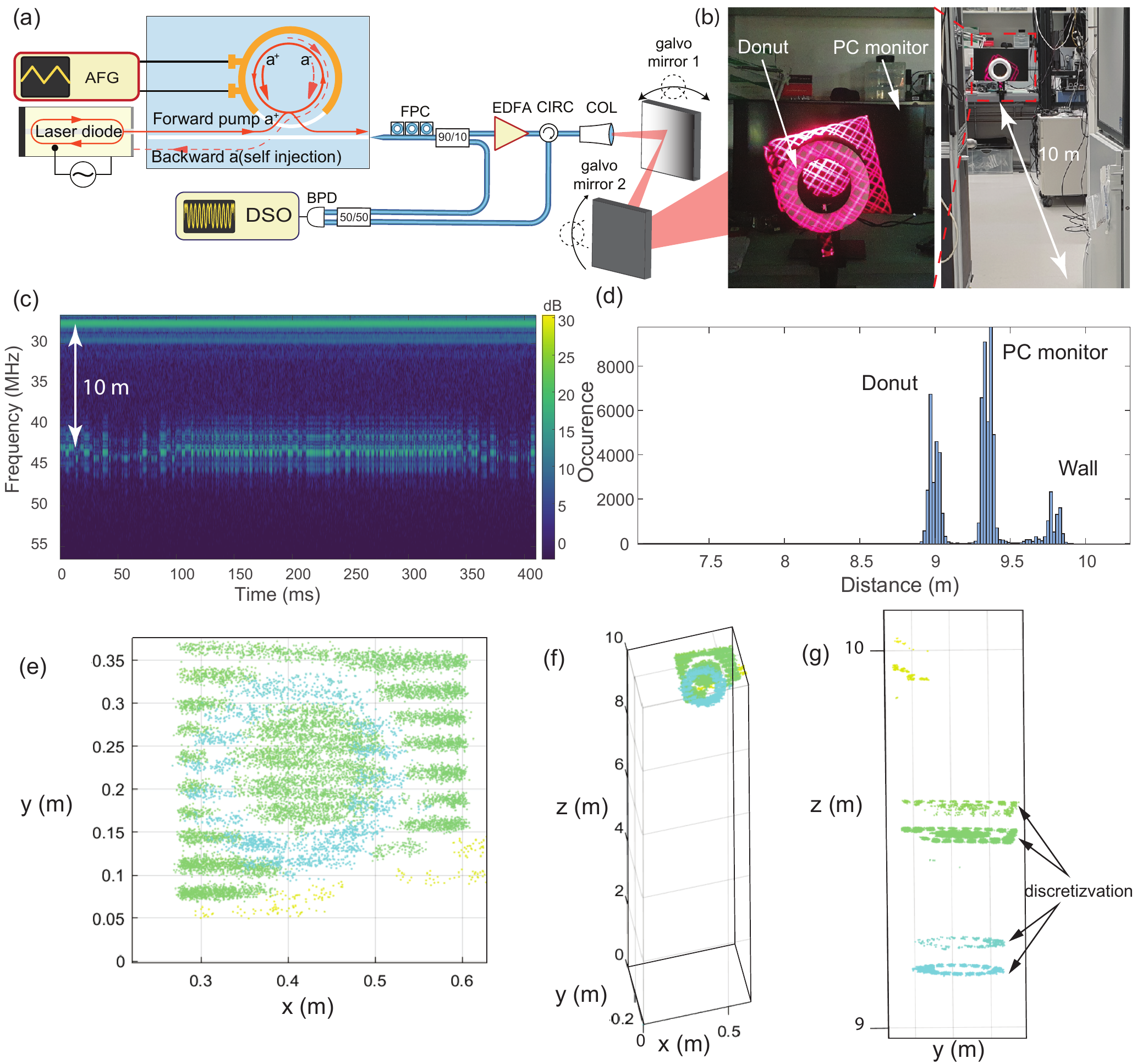}
\caption{
\textbf{Optical ranging using the frequency-agile, hybrid integrated laser.}
(a) Schematic of the setup for FMCW LiDAR measurement.
A triangular ramp with 150~V peak-to-peak amplitude at 100~kHz rate is applied to the AlN piezoactuator providing a 1.2~GHz optical frequency excursion of the self-injection locked DFB laser.
Beam steering is realized using a mechanical galvo scanner with two mirrors.
(b) Photos of the target - a donut in front of a PC monitor.
(c) Time-frequency plot for a signal from the target.
\textcolor{black}{(d) Histogram of distance distribution in the point cloud.}
(e,f,g) Point cloud of the target from different perspectives using a beam scanning pattern with 3~Hz vertical and 60~Hz horizontal triangular scanning frequencies.
Point colors are based on distance. Donut corresponds to blue, PC monitor - green, wall - yellow.
}
\label{Fig:Fig4}
\end{figure*}
%%%%%%%%%%%%%%%%%%%%%%%%%%%%%%%%%%%

\noindent \textbf{Optical FMCW LiDAR using the hybrid integrated laser}.
As an actual demonstration of the potential of the hybrid integrated laser, we perform optical FMCW LiDAR mapping in the laboratory environment.
Importantly, we can - due to the excellent linearity, low hysteresis, and narrow linewidth - perform ranging \emph{without any adaptive clock sampling and without pre-distortion linearization}.
Figure~\ref{Fig:Fig4}(a) shows the experimental setup of FMCW LiDAR measurement, for the description of the experiment refer to Methods.
We apply triangular chirp with 100~kHz frequency to piezoelectric AlN actuator to obtain 1.2~GHz optical frequency excursion of self-injection locked laser, corresponding to \textcolor{black}{12.5~cm} resolution in distance measurement.
Beam steering is realized using mechanical galvo scanner with two mirrors.
For the ranging target scene, we use a polystyrene foam donut in front of a PC monitor 10 m away from the laser collimator (Fig.\ref{Fig:Fig4}(b)).
We record a beat signal of light reflected from the target and the laser in the local oscillator path on a balanced PD.
To construct the point cloud from a recorded oscillogram, we first employ a short-time Fourier transform with a window size equal to half of the chirping period.
Obtained time-frequency plots are presented in Fig.~\ref{Fig:Fig4}(c) for the target and in Fig.~\ref{Fig:Fig4}.
Time-frequency spectrograms contain 82k timeslices with typical $~$15 dB SNR for the target.
We remove points with SNR below 10~dB from the point cloud.
Noticeable reflection at 27~MHz in Fig.~\ref{Fig:Fig4}(c) is due to the reflection from the collimator.
Peaks at 40-46~MHz offsets correspond to the target scene.
\textcolor{black}{We find a peak with maximal spectral amplitude in the time-frequency plot for each timeslice. The frequency of the peak provides the distance information (radial coordinate) for each timeslice.}
\textcolor{black}{Fig.~{\ref{Fig:Fig4}(d)} provides a histogram of distance distribution for the target point cloud. We used zero-padding to increase FFT window size four times and to obtain a continuous distribution of radial distance. Peaks at 9 m distance correspond to the donut, 9.35 m to the PC monitor, and 9.7 m to the wall.}
Polar and azimuthal coordinates were retrieved from the galvo scanner mirrors' driving signals, which were recorded on the same DSO.
See attached to the SI documented interactive code for LiDAR data processing.
Figure~\ref{Fig:Fig4}(e,f,g) shows the point cloud of the scene with distance-based coloring, the donut is depicted in blue, PC monitor in green, and part of the wall in yellow.
The scene has been properly reconstructed, although the dual layering of objects is visible in Fig.~\ref{Fig:Fig4}(g) due to a discretization error from the \textcolor{black}{12.5~cm} LiDAR resolution (no zero-padding was used in point cloud data processing).

To increase optical frequency excursion limited in our method and particular \SiN sample by injection locking range and improve LiDAR resolution, one might consider a different tuning scheme where the laser diode current and the piezoactuator voltage are synchronously tuned in a way that the laser is kept in injection-locked state.
In this case, a diode current tuning might not be precisely linear as self-injection locking implies and preserves the linearity of cavity resonance tuning by the piezoelectrical actuator.
In such a feed-forward scheme, the laser tuning range would be limited by piezoactuator's tuning range. \textcolor{black}{Another option might be to use a two microring-based Vernier configuration for the laser} \cite{vanRees:20,srinivasan2015coupled}, \textcolor{black}{with the fast tuning provided by integrated piezoactuators}.

\begin{figure}[t!]
\centering
\includegraphics[width=0.4\textwidth]{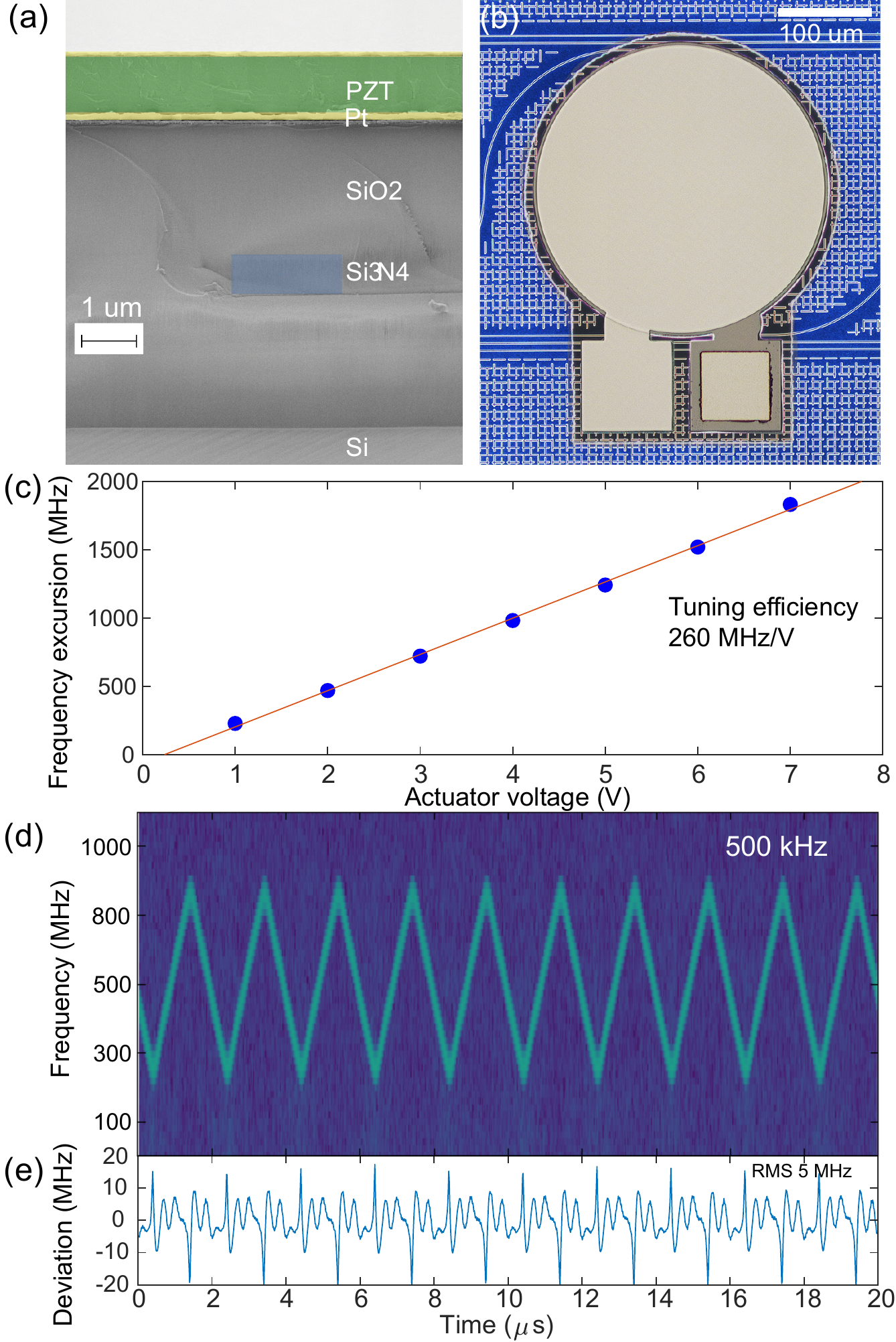}
\caption{
\textcolor{black}{
\textbf{Low voltage fast tuning with integrated PZT piezoactuator.}
(a)~False-colored SEM image of the sample cross-section, showing the PZT actuator integrated on the Si$_3$N$_4$ photonic circuit.
The piezoelectric actuator is composed of Pt (yellow), PZT (green) layers on top of Si$_3$N$_4$ (blue) buried in SiO$_2$ cladding.
(b)~Optical micrograph of disk-shaped PZT actuator on top of \SiN microring with 100 GHz FSR.
(c)~Frequency excursion as a function of the voltage applied to the PZT actuator, measured at 81 kHz triangular chirping rate. Linear fit (red) provides 260 MHz/V tuning efficiency.
(d)~Time-frequency spectrogram of the heterodyne beat-note for 500 kHz triangular chirp repetition frequency with 2.2 V applied to the PZT actuator.
(e)~Residual of least-squares fitting of the time-frequency trace with symmetric triangular chirp pattern.
}
}
\label{Fig:Fig5}
\end{figure}

\textcolor{black}{
\noindent \textbf{Hybrid integrated laser with fast tuning based on integrated low voltage PZT piezoactuator.}
We investigate the reduction of tuning voltage requirement by introducing an integrated piezoelectric actuator based on lead zirconium titanate (PZT) \cite{Hosseini:15}. The actuator has PZT as the main piezoelectric material and platinum (Pt) as the top and the bottom (ground) electrodes, as shown in Fig.~\ref{Fig:Fig5} (a). The key difference to AlN is that the PZT process has a patterned ground plane to eliminate bond-pad capacitance, since the relative permittivity of PZT is >800 compared to $\sim9$ for AlN. Using disk-shaped PZT actuator on top of \SiN microresonator with 100 GHz FSR (see Fig.~\ref{Fig:Fig5} (b)), we perform the heterodyne beat experiment with fixed frequency ECDL reference laser. The DFB is self-injection locked at 240 mA driving current to the \SiN cavity resonance. By applying a triangular voltage ramp with an 81 kHz frequency and an amplitude from 1 V to 7 V and positive bias of 7V, we measured optical frequency excursion from 230 MHz to 1832 MHz correspondingly, resulting in 260 MHz/V tuning efficiency (see Fig.~\ref{Fig:Fig5} (c)). Figure ~\ref{Fig:Fig5} (d,e) presents the laser frequency spectrogram and the corresponding tuning nonlinearity at 500 kHz ramp frequency with 2.2 V applied, confirming the PZT actuator nonlinearity 0.95\% over 525 MHz frequency excursion. With the PZT breakdown voltage of 25 V we expect the maximal $\sim6$ GHz optical tuning range using the feed-forward method.
}

%%%%%%%%%%%%%%%%%%%%%%%%%%%%%%%%%%%%%%%%%%%%%%%%%%%%%%%%%%%%%%%%%%%%%%%%%%%%%%%%
\section*{Conclusion}
We have demonstrated a hybrid integrated laser with ultralow noise - on par with a fibre laser - while exhibiting unsurpassed flat and high frequency actuation response.
The approach is based on foundry-ready processes that include photonic integrated circuits based on \SiN as well as AlN \textcolor{black}{and PZT} MEMS processing, and is therefore amenable to CMOS compatible large-volume manufacturing.
The system is ideally suited for medium to long-range LiDAR, as required for autonomous driving, drone navigation, or industrial and terrain mapping.
%While the hybrid integrated laser system demonstrated in this work is able to substantially reduce the linewidth of the semiconductor laser diode and \textcolor{black}{TJK: I thought we leave out the DKS?}generate DKS microcomb, the frequency agility that is enabled by the high-bandwidth AlN actuator brings the system into a realm of important applications that could not be realised before. Medium to long-range FMCW LiDAR that requires highly linear chirped optical signals with high spectral purity and a large chirping range is difficult to achieve with integrated laser systems. Using the hybrid integrated laser system we developed here, we show LiDAR demo in the lab environment. %One can expect in the near future that FMCW LiDAR suitable for an autonomous vehicle may be materialized using demonstrated self-injection locking principle.
With our demonstration, many other applications, which have traditionally relied on external cavity or fibre lasers, such as distributed fibre sensing, coherent communications with large spectral efficiency, or spectroscopy, may equally benefit from the low-noise and fast tunable lasers.
In addition, while our lasers were demonstrated at 1556~nm, the center wavelength can be readily extended to other ranges, including the near IR and mid-infrared, due to the transparency of Si$_3$N$_4$.
\textcolor{black}{To further improve the performance of the laser system here, one can use laser chips with higher output power. Also, microresonators with larger mode volumes can be used to reduce the fundamental thermo-refractive noise that limits the intrinsic laser linewidth.}
%Full packaging of the system that stabilizes both the temperature and coupling gap will substantially mitigate the frequency instability at low frequency ranges.
In addition, with careful design of the geometry of the piezoactuator and the dimension of the \SiN photonic chip to suppress HBAR modes, the linear actuation bandwidth could extend into the GHz regime - limited only by the internal modes of the piezoactuator \cite{tian2020hybrid}.
Last, viewed more broadly, our results are an intriguing example where hybrid integrated photonics yields not only substantial reductions in size and weight of narrow-linewidth frequency-agile lasers, but in addition, unprecedented performance metrics in terms of frequency agility and phase noise, superseding the workhorse external-cavity diode lasers, and on par with fibre lasers.
Our results indicate the significant potential of hybrid integrated photonics to replace decades-old prevailing technologies.

\noindent \textbf{Methods}

\medskip
\begin{footnotesize}
%\noindent \textbf{Cavity ringdown}:
%
%\noindent \textbf{Thermal simulations}:
%
%\noindent \textbf{Heterodyne characterisation of frequency-modulation}:

\noindent \textbf{Laser frequency noise measurement.}
The laser frequency noise is measured using two methods: First, we perform heterodyne beat note spectroscopy using an external cavity diode reference laser (Toptica CTL), which we lock to a crystalline whispering gallery mode microresonator via the Pound-Drever-Hall (PDH) technique to suppress characteristic phase noise peaks of the diode laser between 1 and 10~kHz. The noise of the reference laser is determined by heterodyne spectroscopy with a fibre laser locked to an ultra-stable optical cavity (MenloSystems ORNS). \textcolor{black}{ Direct comparison of the ultrastable reference at 1552.5 nm is not possible due to a large wavelength offset of DFB, operating at 1556 nm at 26 C temperature.} The electrical output of the photodiode is fed to a phase noise analyzer (Rhode \& Schwarz FSW43). The phase noise of the lasers is determined via Welch's method from a time sampling trace of the in-phase and quadrature components of the beat note and transformed according to $S_{ff} = f^2 \cdot S_{\phi\phi}$ to yield the laser frequency noise. At high frequencies, we observe that the direct heterodyne measurement of the noise is either limited by photon shot noise above 1~MHz, the noise of the reference lasers or the PDH locking bandwidth.

\noindent The second method by which we measure the frequency noise of the SIL laser is cross-correlation spectroscopy. Here, we forgo the PDH locking of the reference laser and instead use a second similar reference laser to record another beat note. The reference lasers are tuned to decrease both beating frequencies to 10~MHz, and the signals are sampled directly on a fast sampling digitizer with 50~Msps for 100~s. The phase of both beat notes is determined by the Hilbert transform of the sampling trace. The unwrapped phase is dechirped by linear regression, and the phase noise spectrum is determined by Fourier transform. Details of the cross-correlation method are given in references \cite{rubiola2000correlation,xie2017phase}. The true phase noise of the SIL laser is revealed after averaging of the phase noise cross-spectral power density over many time intervals $m$ according to $S_{\phi\phi} (f) = \Re{\langle S_{AB} (f) \rangle}_{m} + \mathcal{O}\left( m^{-1/2} \right) $, where the subscripts $A,B$ denote the two heterodyne beat notes and $S_{AB}$ their cross spectral density. The biggest drawback of the cross-correlation measurement comes from the requirement of exceedingly long averaging times for the determination of the noise at low offset frequencies. All the relevant information on the device phase noise is retained in the real part of the cross-power spectral density $\Re{\langle S_{AB}\rangle}$, and the correlated phase noise of the SIL laser appears as a positive correlation only \cite{gruson2020artifacts}. By analyzing its sign, we can identify the low frequency cut-off below which averaged cross power spectral density is dominated by residual random (anti-)correlation of the ECDL noises. We find that below 5~kHz the unlocked ECDL noise becomes too large and completely dominates the SIL laser noise in our measurement. The frequency flicker noise floor of both measurement methods agrees well in the region between 10~kHz and 1~MHz, where the cross correlation method reveals several noise peaks of the reference laser.

\noindent To calculate TRN limit, we use the following expression for the effective temperature fluctuations \cite{huang2019thermorefractive, kondratiev2018thermorefractive}:

\begin{equation*}
	S_{\delta T}(\omega) = \dfrac{k_B T^2}{\sqrt{\pi^3 \kappa \rho C \omega}}\sqrt{\dfrac{1}{2p+1}}\dfrac{1}{R\sqrt{d_r^2-d_z^2}}\dfrac{1}{[1+(\omega\tau_d)^{3/4}]^2},
\end{equation*}

\noindent where $R$ is the microring resonator radius; \textcolor{black}{\SiN material parameters} $\rho=3.29\times 10^3$ kg$\cdot$m$^{-3}$ is density; $\kappa=30$ W$\cdot$m$^{-1}$$\cdot$K$^{-1}$ is thermal conductivity; $C=800$ J$\cdot$kg$^{-1}$$\cdot$K$^{-1}$ is specific heat capacity; $T=300$ K, $d_z=1.5$ $\mu$m and $d_r=0.75$ $\mu$m stand for halfwidths of the fundamental mode, with orbital number $l$, azimuthal number $m$ and meridional mode number $p=l-m$, $\tau_d = \dfrac{\pi^{1/3}}{4^{1/3}}\dfrac{\rho C}{\kappa}d_r^2$.

\noindent \textbf{FMCW laser ranging experiment.} The DFB is directly butt-coupled to \textcolor{black}{a 190.7-GHz FSR microresonator on a regular non-apodized chip placed on a carbon tape. The chip features a single AlN disk actuator.} We employ laser frequency tuning by keeping the laser diode's current fixed, the laser self-injection locked to a \SiN resonance and by tuning only the cavity resonance with AlN piezoactuator. Triangular ramp with 100~kHz frequency from an arbitrary waveform generator is amplified to 150~V (peak-to-peak amplitude) with a high voltage amplifier with 5~MHz bandwidth. The diode current is set near the center of the injection locking range (281~mA for the particular resonance used) and AlN voltage is adjusted to keep the laser inside the self-injection locking range, resulting in a 1.2-GHz optical frequency excursion, corresponding to a FMCW LiDAR resolution of approximately \textcolor{black}{12.5~cm}. While not required in actual operation, a small portion of the laser power was split of (95/5) to a fibre-coupled Mach-Zehnder interferometer for calibration of the frequency excursion. The length of the calibration MZI was measured with a frequency comb calibrated tunable diode laser scan. No active linearization or k-point sampling was necessary due to the excellent linearity and negligible hysteresis of the integrated AlN tuner. While 10$\%$ of the total laser power of 1.5~mW is used as a local oscillator (LO), 90$\%$ of the light is amplified by an EDFA to 10~mW and sent to a target through a collimator with an aperture of 8~mm, which was adjusted to fit the target range of 10~m. A double-axis galvanometric mirror scanner (Thorlabs GVS112) was used for the beam steering. \textcolor{black}{Vertical and horizontal mirrors were rotated with constant speed at rates 3 Hz and 60 Hz correspondingly to cover the full scene during the measurement time of 400 ms.}
%The coherent LiDAR measurement was performed using the 192~GHz FSR ring with monolithically integrated AlN actuators for self-injection locking. The fibre-coupled aoutput power was 1.5~mW and amplified to 8~mW for the measurement using an erbium-doped fibre amplifier. The main portion was split into the signal and reference arms (95/5) of the main ranging interferometer. The signal was emitted from a 8~mm fibre coupled zoom collimator, which was adjusted to fit the target range of 10~m. A double-axis galanometric mirror scanner was used to

\noindent \textbf{Funding Information}: This work was supported by funding from the European Union H2020 research and innovation programme under FET-Open grant agreement no. 863322 (TeraSlice) and the Marie Sklodowska-Curie IF grant agreement No. 846737 (CoSiLiS). This material is based upon work supported by the Air Force Office of Scientific Research under award number FA9550-19-1-0250. This work was supported by funding from the Swiss National Science Foundation under grant agreement No.176563. This work was supported by funding from the European Union H2020 research and innovation programme under HOT:  FET-Proactive grant agreement No. 732894. This work was also supported by the United States’ National Science Foundation under Award 18-39164 and QIS DCL 20-063.  The chip samples were fabricated in the EPFL center of MicroNanoTechnology (CMi), and in the Birck Nanotechnology Center at Purdue University. AlN deposition was performed at OEM Group Inc. PZT layer was fabricated at Radiant Technologies Inc.

\noindent \textbf{Author contribution}:
G.L and J.R performed the experiments with the help from V.S.
W.W. developed a theoretical model and performed numerical simulations.
A.S. performed mechanical response measurements.
J.L., H.T., R.N.W. and J.H. designed, fabricated, and characterized the \SiN chips.
R.N.W. performed chip apodization.
G.L, J.R., A.S., V.S analyzed the data.
W.W., G.L., J.R. wrote the manuscript with input from all authors.
T.J.K. and S.A.B. supervised the project.

\noindent \textbf{Acknowledgments}:
We acknowledge Lin Chang and John Bowers for providing the DFB laser.
%We thank Freedom Photonics for providing the DFB laser.

\noindent \textbf{Data Availability Statement}: The code and data used to produce the plots within this work will be released on the repository \texttt{Zenodo} upon publication of this preprint.

\noindent\textbf{Correspondence and requests for materials} should be addressed to T.J.K.

%\noindent\textbf{Note added:} During the preparation of this work we became aware of work of K.Vahala and J.Bowers groups reporting on fixed frequency ultra narrow linewidth lasers \cite{jin2020hertzlinewidth}.

\end{footnotesize}
\bibliography{citation}

\end{document}

\title{Supplementary Information for\\
	Ultralow-noise frequency-agile photonic integrated lasers}

\author{Grigory Lihachev}
\email{These authors contributed equally}
\affiliation{Institute of Physics, Swiss Federal Institute of Technology Lausanne (EPFL), CH-1015 Lausanne, Switzerland}

\author{Johann Riemensberger}
\email{These authors contributed equally}
\affiliation{Institute of Physics, Swiss Federal Institute of Technology Lausanne (EPFL), CH-1015 Lausanne, Switzerland}

\author{Wenle Weng}
\email{These authors contributed equally}
\affiliation{Institute of Physics, Swiss Federal Institute of Technology Lausanne (EPFL), CH-1015 Lausanne, Switzerland}

\author{Junqiu Liu}
\affiliation{Institute of Physics, Swiss Federal Institute of Technology Lausanne (EPFL), CH-1015 Lausanne, Switzerland}

\author{Hao Tian}
\affiliation{OxideMEMS Lab, Purdue University, 47907 West Lafayette, IN, USA}

\author{Anat Siddharth}
\affiliation{Institute of Physics, Swiss Federal Institute of Technology Lausanne (EPFL), CH-1015 Lausanne, Switzerland}

\author{Viacheslav Snigirev}
\affiliation{Institute of Physics, Swiss Federal Institute of Technology Lausanne (EPFL), CH-1015 Lausanne, Switzerland}

\author{Rui Ning Wang}
\affiliation{Institute of Physics, Swiss Federal Institute of Technology Lausanne (EPFL), CH-1015 Lausanne, Switzerland}

\author{Jijun He}
\affiliation{Institute of Physics, Swiss Federal Institute of Technology Lausanne (EPFL), CH-1015 Lausanne, Switzerland}

\author{Sunil A. Bhave}
\affiliation{OxideMEMS Lab, Purdue University, 47907 West Lafayette, IN, USA}

\author{Tobias J. Kippenberg}
\email[]{tobias.kippenberg@epfl.ch}
\affiliation{Institute of Physics, Swiss Federal Institute of Technology Lausanne (EPFL), CH-1015 Lausanne, Switzerland}

\maketitle

% \renewcommand\thefigure{S.\arabic{figure}}
% \renewcommand\theequation{S.\arabic{equation}}
% \setcounter{figure}{0}
% \setcounter{equation}{0}

%%%%%%%%%%%%%%%%%%%%%%%%%%%%%%
\subsection*{Simulation of the self-injection locking dynamics}
%%%%%%%%%%%%%%%%%%%%%%%%%%%%%%

The conventional laser rate equation \cite{agrawal1984line} for describing the semiconductor laser field dynamics can be written as:
%%%%%%%%%%%%%%%%%%%%%%%%%%%%%%
\begin{equation}
\label{laserfield}
\frac{dA_\mathrm{laser}}{dt} = \frac{1}{2} (1 - i \alpha_\mathrm{H}) [a V (N-N_0) - \frac{1}{\tau_\mathrm{p}}]A_\mathrm{laser} + F_\mathrm{A}
\end{equation}
%%%%%%%%%%%%%%%%%%%%%%%%%%%%%%
where $A_\mathrm{laser}$ is the complex laser field profile and $|A_\mathrm{laser}|^2$ is the averaged photon density in the laser cavity, $N$ is the carrier density, $\alpha_\mathrm{H}$ is the linewidth enhancement factor, $a$ is the differential gain, $V$ is the laser active volume, $N_0$ is the carrier density at transparency, and $F_\mathrm{A}$ is the Langevin noise term to describe the spontaneous emission into the lasing field \cite{dellunde1997frequency}. This Gaussian noise term obeys $\left<F_\mathrm{A}(t)F_\mathrm{A}^*(t')\right> = 4 \beta N B_\mathrm{sim} \delta(t-t')$. Here $\beta$ is a factor related to the amount of spontaneous emission into the lasing mode, $B_\mathrm{sim}$ is the simulation bandwidth. We note that the noise in the carrier density fluctuation is ignored in our simulations because its contribution to the laser noise is insignificant in comparison to the noise in the optical field \cite{ohtsubo2012semiconductor}.

%%%%%%%%%%%%%%%%%%%%%%%%%%%%%%%%%%%%%%%%%
%%%%%%%%%%%%%%%%%%%%%%%%%%%%%%%%%%%%%%%%%
\begin{table}[htbp]
\centering
%\caption{Values and definitions of parameters used in simulations.}
\begin{tabular}{|c|c|c|c|}
\hline
Symbol & Value & Unit & Definition \\
\hline
$\alpha_\mathrm{H}$ & $5$   &                 & Linewidth enhancement factor \\
$a$ & $1\times10^{4}$   & s$^{-1}$       & Differential gain \\
$N_0$ & $1\times10^{24}$   & m$^{-3}$                & Carrier density at transparency \\
$\kappa$ & $1\times10^{11}\times2\pi$   & rad s$^{-1}$         & Laser cavity loss rate\\
$I_\mathrm{bias}$ & $250$   & mA                & Biased current \\
$\zeta$ & $5\times10^{8}$   & Hz/mA                & Current-frequency tuning coefficient \\
$\gamma$ & $1\times10^9$         & s$^{-1}$           & Carrier recombination rate\\
$V$ & $2\times10^{-16}$   & m$^{3}$                & Volume of active section \\
$\beta$ & $1\times10^{4}$   & s$^{-1}$    & Spontaneous emission coefficient \\
$e$ & $1.6\times10^{-19}$   & C                & Elementary electronic charge \\
%\hline
$\kappa_\mathrm{r}$ & $60\times10^6\times2\pi$         &  rad s$^{-1}$      & Loaded loss rate \\
$\kappa_\mathrm{ex}$ & $50\times10^6\times2\pi$     &  rad s$^{-1}$      & External coupling rate \\
$\kappa_\mathrm{sc}$ & $30\times10^6\times2\pi$         &  rad s$^{-1}$      & CW-CCW coupling rate \\
$\kappa_\mathrm{inj}$ & $7.5\times10^{16}\times(2\pi)^2$         &  rad$^2$ s$^{-2}$      & Laser-microresonator coupling factor \\
$\phi$ &    $1.5$         &  rad      & Feedback phase \\
$n_\mathrm{2}$ & $2.4\times10^{-19}$         &  m$^2$/W      & Kerr nonlinear index \\
$n_0$ & $2$         &        & Microresonator refractive index \\
\hline
\end{tabular}
\caption{Values and definitions of parameters used in the simulations.}
  \label{table1}
\end{table}
%%%%%%%%%%%%%%%%%%%%%%%%%%%%%%%%%%%%%%%%%
%%%%%%%%%%%%%%%%%%%%%%%%%%%%%%%%%%%%%%%%%

We incorporate the laser rate equations with the equations for the coupled fields in a clockwise (CW) mode (which is the mode directly pumped by the laser field) and the counter-clockwise (CCW) mode in the microresonator to numerically study the self-injection locking dynamics. We also add a phenomenological term in the laser field equation in order to include the current-frequency tuning ability with a coefficient of 500~MHz/mA. This frequency tuning ability originates from several mechanisms include gain compression, carrier-density-dependent refractive index change and temperature change \cite{welford1985magnitude,ohtsubo2012semiconductor}. Since only $\sim10$~mA is applied for the current modulation amplitude, the linear approximation agrees well with the experimental observations. The coupled equations are written as:
%%%%%%%%%%%%%%%%%%%%%%%%%%%%%%
\begin{equation}
	\label{eq1}
	\frac{dN}{dt} = \frac{I_\mathrm{bias} + \Delta I}{eV} - \gamma N - aV(N-N_0) |A_\mathrm{laser}|^2
\end{equation}
%%%%%%%%%%%%%%%%%%%%%%%%%%%%%%
\begin{equation}
	\label{eq2}
	\frac{dA_\mathrm{laser}}{dt} = \left[\frac{1}{2} (1 - i \alpha_\mathrm{H}) (a V (N-N_0) - \kappa) -i\zeta\Delta I \right]A_\mathrm{laser} + \sqrt{\kappa_\mathrm{inj}} e^{i \phi} A_\mathrm{CCW} + F_\mathrm{A}
\end{equation}
%%%%%%%%%%%%%%%%%%%%%%%%%%%%%%
\begin{equation}
	\label{eq3}
	\frac{dA_\mathrm{CW}}{dt} = (-\kappa_\mathrm{r} + i \delta\omega(t) + i \Gamma (|A_\mathrm{CW}|^2) + 2 |A_\mathrm{CCW}|^2) A_\mathrm{CW} + i \kappa_\mathrm{sc} A_\mathrm{CCW} + \sqrt{\kappa_\mathrm{inj}} e^{i \phi} A_\mathrm{laser}
\end{equation}
%%%%%%%%%%%%%%%%%%%%%%%%%%%%%%
\begin{equation}
	\label{eq4}
	\frac{dA_\mathrm{CCW}}{dt} = (-\kappa_\mathrm{r} + i \delta\omega(t) + i \Gamma (|A_\mathrm{CCW}|^2) + 2 |A_\mathrm{CW}|^2) A_\mathrm{CCW} + i \kappa_\mathrm{sc}^* A_\mathrm{CW}
\end{equation}
%%%%%%%%%%%%%%%%%%%%%%%%%%%%%%

Here the carrier recombination rate is represented by $\gamma$, the elementary charge is denoted by $e$, $|A_\mathrm{CW}|^2$ and $|A_\mathrm{CCW}|^2$ are the averaged photon densities in the CW microresonator mode and the CCW microresonator mode respectively, $\kappa_\mathrm{r}$ is the microresonator decay rate, $\kappa_\mathrm{sc}$ is the coupling rate between the CW and the CCW modes, $\kappa_\mathrm{inj}$ is the coupling rate between the microresonator and the laser, and $\phi$ is the coupling phase related to the gap between the laser and the microresonator. The Kerr frequency shift coefficient can be calculated using $\Gamma = \frac {\hbar \omega_0^2 c n_2}{n_0^2}$, where $n_0$ is the refractive index of the microresonator material and $n_2$ is the Kerr nonlinear coefficient. The detailed definitions and values of the parameters used in the simulations are presented in Table\,\ref{table1}.

%%%%%%%%%%%%%%%%%%%%%%%%%%%%%%%%%%%%%%%%%
\begin{figure*}[hbt]
\centering
%\hspace{20mm}
\includegraphics[width=0.7\textwidth]{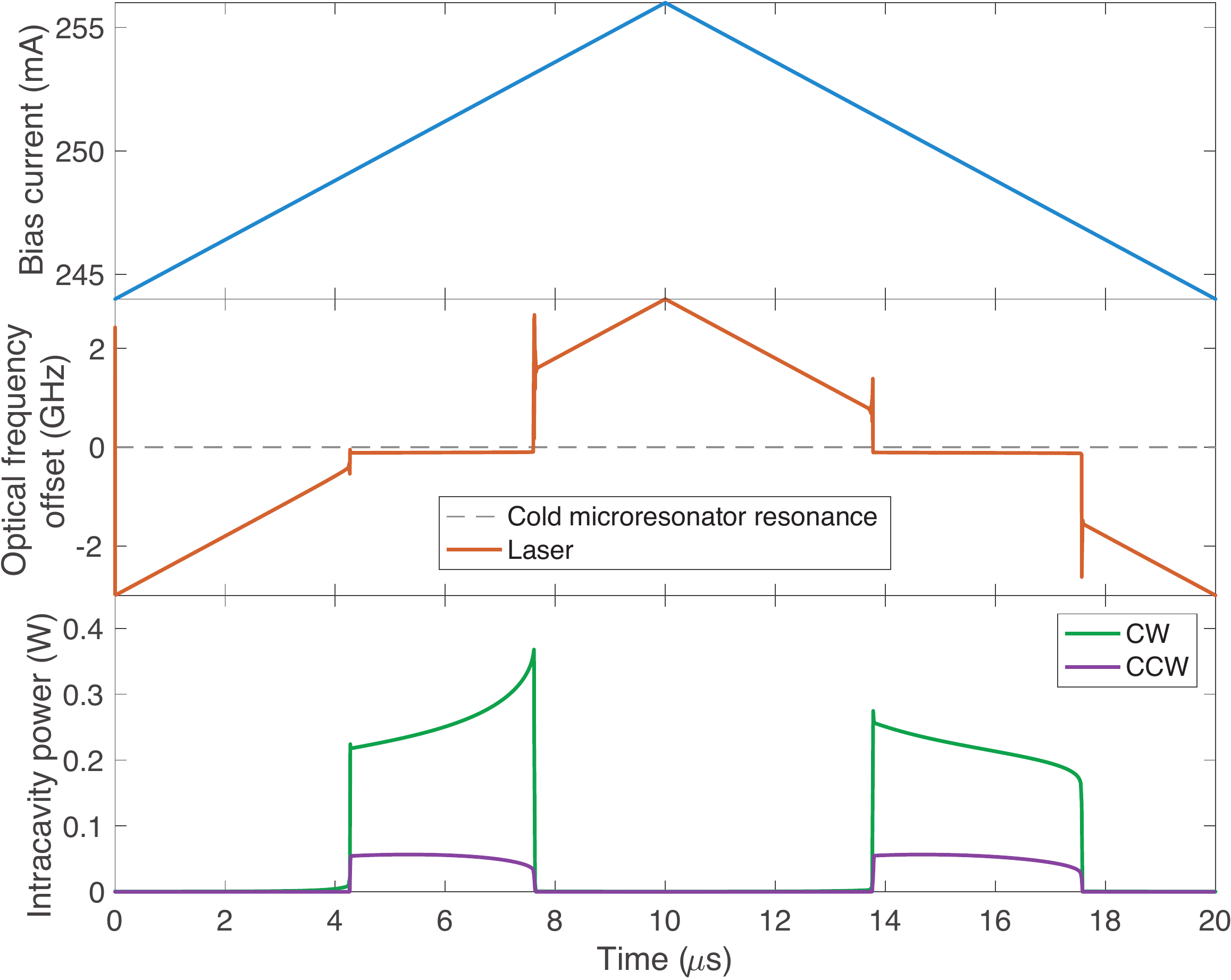}
\caption{\textbf{Simulated self-injection locking with laser current tuning.} The bias current is tuned forward and backward with a total tuning range of 12~mA, corresponding to a natural lasing frequency range of 6~GHz. The middle panel presents the frequency shift of the laser, showing qualitative agreement with the experimental result in Fig.\,1\,(c) in the main text. In the injection-locked ranges, the small frequency gap between the cold microresonator resonance and the lasing frequency is caused by the Kerr-effect-induced frequency shift of the microresonator resonances. In the lower panel, the intracavity powers of the CW and CCW fields are plotted.
}\label{figS1}
\end{figure*}
%%%%%%%%%%%%%%%%%%%%%%%%%%%%%%%%%%%%%%%%%
%%%%%%%%%%%%%%%%%%%%%%%%%%%%%%%%%%%%%%%%%

Fig. \ref{figS1} shows the simulation results when the laser bias current is swept over 12~mA. The laser frequency shows the self-injection-locking phenomenon at both current-up and -down sweeping directions. We adjust the feedback phase $\phi$, so the locking ranges with different sweeping directions are similar. Prior studies show that a well-selected phase facilitates the maximization of the locking range \cite{kondratiev2017self}. The simulated locking range of a few gigahertz is in good agreement with the experimental results presented in Fig.\,1\,(c) in the main text.

%%%%%%%%%%%%%%%%%%%%%%%%%%%%%%%%%%%%%%%%%
\begin{figure*}[hbt]
\centering
\includegraphics[width=0.95\textwidth]{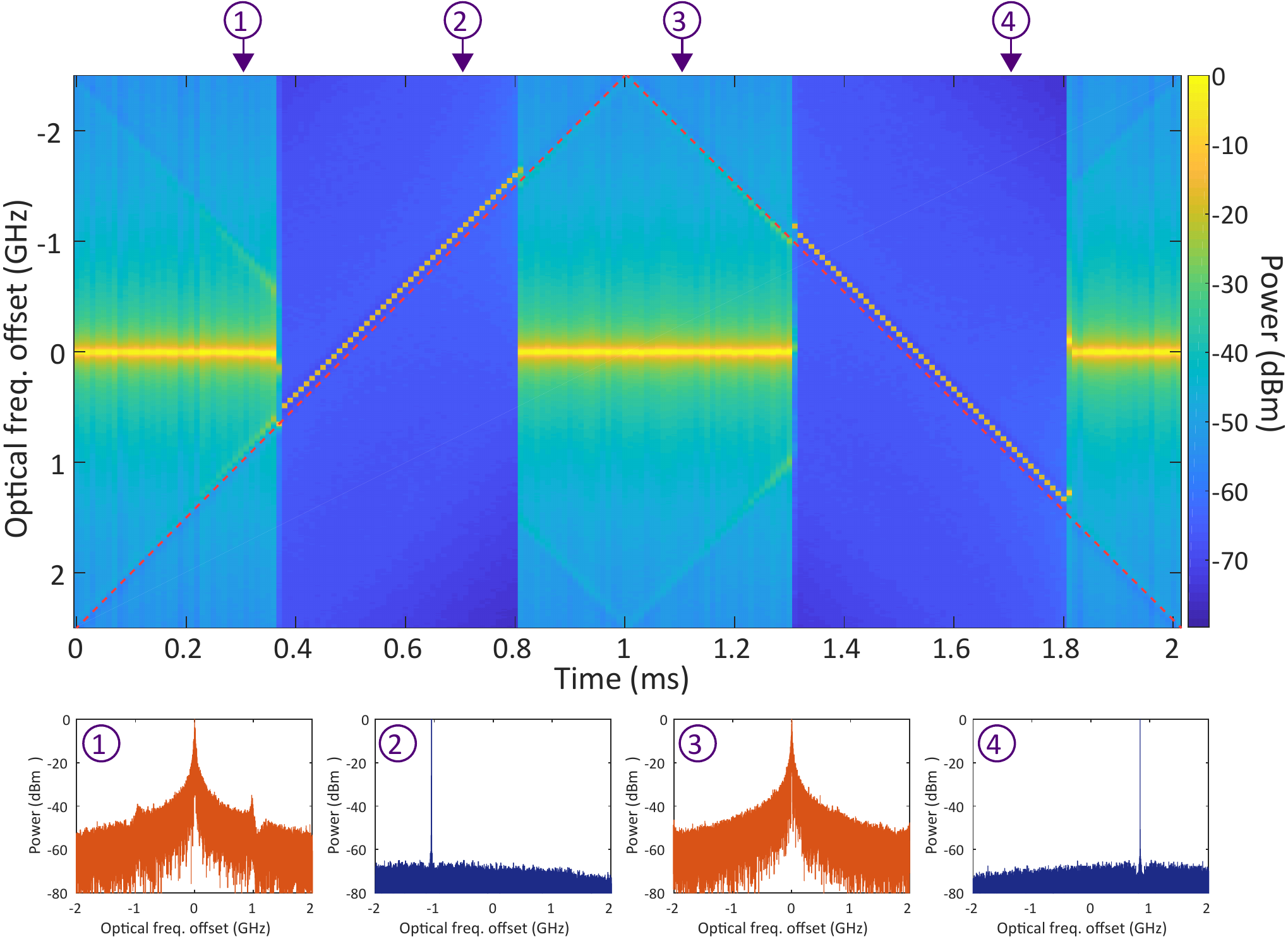}
\caption{\textbf{Simulated laser frequency spectrogram as the microresonator resonance is swept back and forth.} Each frame in the spectrogram is normalised, so the peak power is 0~dBm. The red dashed lines show the cold microresonator resonance frequencies. Four spectra labeled with 1 -- 4 are shown at the bottom, corresponding to the numbered positions indicated in the spectrogram. The spectra of the self-injection-locked laser (in blue) show significantly narrower lasing linewidth and lower noise floor than those of the free-running laser (in red).
}\label{spectrogram}
\end{figure*}
%%%%%%%%%%%%%%%%%%%%%%%%%%%%%%%%%%%%%%%%%

Next, we sweep the cold microresonator resonance frequency over the semiconductor laser's natural lasing frequency with both frequency-up and frequency-down directions while the laser bias current is kept constant. We use fast Fourier transform to compute the laser optical spectra at varied laser-microresonator detunings. Fig. \ref{spectrogram} shows the full spectrogram. The evolution of the lasing frequency is similar to the results in \cite{kondratiev2017self}. Additionally, here with the inclusion of the spontaneous emission noise in our simulations, the significant improvement of the laser noise performance of the self-injection locking state is clearly shown.

%%%%%%%%%%%%%%%%%%%%%%%%%%%%%%%%%%%%%%%%%
\begin{figure*}[hbt]
\centering
\includegraphics[width=0.7\textwidth]{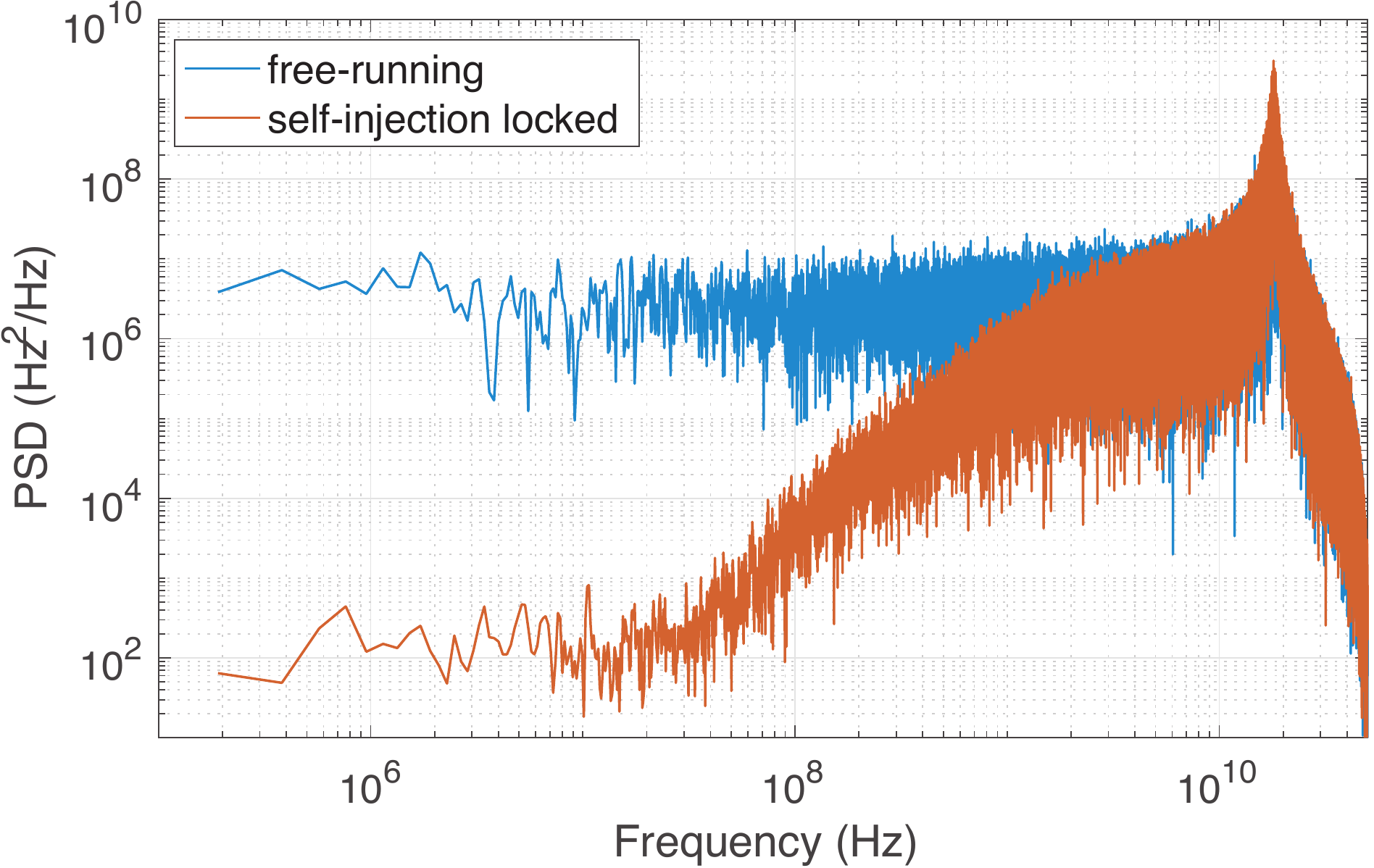}
\caption{\textbf{Simulated laser frequency PSDs.} Within the microresonator bandwidth (i.\,e., 60~MHz), the linewidth narrowing factor is more than $3\times10^4$. The peaks at $\sim20$~GHz are the relaxation oscillations.
}\label{figS3}
\end{figure*}
%%%%%%%%%%%%%%%%%%%%%%%%%%%%%%%%%%%%%%%%%

We compute the laser frequency noise power spectral density (PSD) with the simulation data to characterize the laser linewidth narrowing effect  \cite{laurent1989frequency}. In Fig. \ref{figS3} the PSDs of the free-running (without self-injection) laser and the self-injection-locked laser are compared. This comparison indicates that within the resonance bandwidth of the microresonator the linewidth narrowing factor is larger than 40~dB, showing excellent agreement with the experimental results presented in Fig.\,2\,(d) in the main text.

%%%%%%%%%%%%%%%%%%%%%%%%%%%%%%%%%%%%%%%%%
\begin{figure*}[hbt!]
\centering
\includegraphics[width=0.9\textwidth]{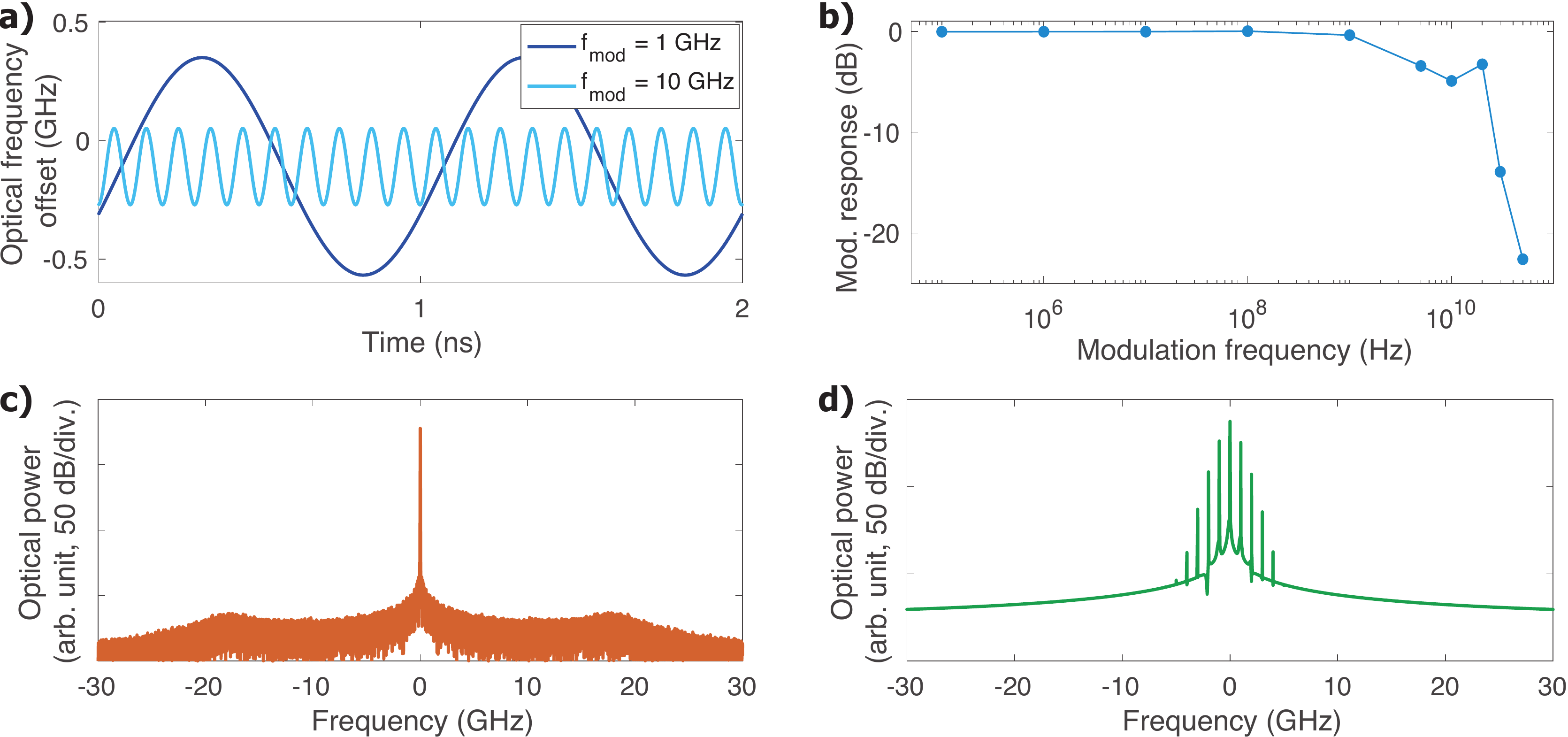}
\caption{\textbf{Simulated microresonator-resonance-modulation responses of the self-injection-locked laser frequency.} \textbf{(a)} With the cold microresonator resonances being modulated with a modulation range of 1~GHz at two different modulation frequencies ($f_\mathrm{mod} =$ 1 and 10~GHz) respectively, the self-injection-locked laser frequency responses are presented. \textbf{(b)} The transfer function of the modulation response. The rise at $f_\mathrm{mod} =$ 20~GHz is due to the nearby relaxation oscillation frequency at 17.8~GHz. \textbf{(c)} Simulated optical spectrum of the free-running laser. \textbf{(d)} Simulated optical spectrum of the intracavity CW field when the microresonator is driven by the free-running laser and resonance-modulated with $f_\mathrm{mod} =$ 1~GHz.
}\label{figS4}
%\vspace{-4mm}
\end{figure*}
%%%%%%%%%%%%%%%%%%%%%%%%%%%%%%%%%%%%%%%%%

To investigate the frequency modulation bandwidth of the hybrid laser, we fix the laser bias current and modulate the cold microresonator resonances with a sine function whose modulation amplitude is of 1~GHz. The modulation frequency ($f_\mathrm{mod}$) is varied from 100~kHz to 50~GHz. The simulated responses of the self-injection-locked laser frequency are displayed in Fig. \ref{figS4} (a) and (b), showing that the laser frequency response is not limited by the resonance bandwidth of the microresonator that provides the feedback for self-injection locking. This result may seem counterintuitive at first. However, our simulations show that when the resonance-modulated microresonator is pumped by a free-running laser, the strong resonance modulation will produce sidebands in the intracavity fields, even when the modulation frequency is much higher than the resonance bandwidth (see Fig. \ref{figS4} (c) and (d)). The created sidebands are fed back to the laser, introducing the amplification of the sidebands. As a result, as long as the sidebands are within the laser cavity resonance bandwidth, the laser may be effectively frequency-modulated while self-injection locked. Of course, in practice, the nonlinear response and the dispersion of the laser gain medium could have a negative impact on the feedback locking dynamics, and in reality, most of the time, the modulation actuation mechanism, i.\,e., the piezoelectric actuation in this work poses the bottleneck of the modulation bandwidth. Yet, our simulation shows the promise for ultrahigh modulation bandwidth for lasers that are self-injection locked with ultrahigh-$Q$ microresonators.

\subsection*{Fabrication of photonic chips with integrated AlN piezoactuators}

Microresonators and bus waveguides are fabricated with the photonic Damascene process \cite{pfeiffer2018photonic}, deep-ultraviolet stepper lithography\cite{liu2020photonic} and silica preform reflow\cite{pfeiffer2018ultra}. The waveguide cross-sections were optimized to minimize field overlap with the waveguide sidewalls and scattering loss. The bus waveguides have the same cross-section as the microring resonators and are terminated at the chip facets with 250~nm wide double-inverse tapers \cite{liu2018double} both for out-coupling to optical fiber and interfacing with the DFB laser. The input coupling loss is 7 dB and can be improved by using dual layer Si$_3$N$_4$ to match the optical mode of a DFB laser.

To fabricate the piezoelectrical actuator, 100~nm Mo and 1 $\mu$m polycrystalline AlN films are sputtered on SiO$_2$ cladding through foundry services (OEM Group). The AlN disk is first patterned and dry-etched using Reactive Ion Etching (RIE) with chemicals Cl$_2$ and BCl$_3$. The dry etching of the bottom electrode (Mo) is performed using Cl$_2$. Finally, the top 100~nm of Al is evaporated and patterned using a standard lift-off process.

\begin{figure*}[htbp]
	\centering
	\includegraphics[width=\columnwidth]{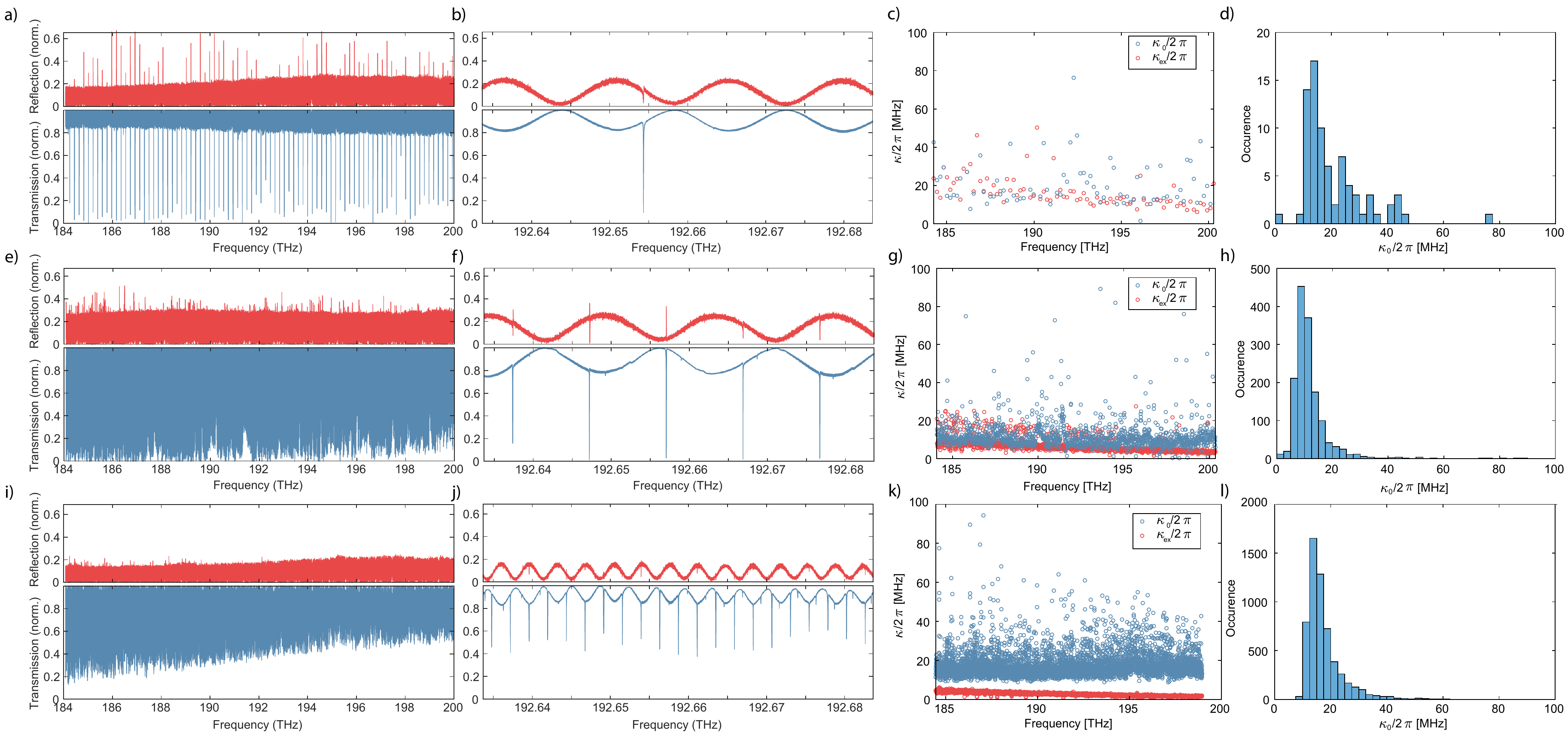}
	\caption{
	\textbf{Linear spectroscopy of photonic components} a,b) Transmission (blue) and reflection (red) of bus waveguide coupled to photonic Damascene microresonator with free-spectral range (FSR) 190.7~GHz. c) Frequency-dependent microresontor loss $\kappa_0$ \textcolor{black}{(blue)} and bus waveguide coupling $\kappa_{ex}$ \textcolor{black}{(red)}. d) Histogram of microresontor loss values. e-h) Same as a-d) but for 9.87~GHz FSR microresonator. i-l) Same as a-d) but for 2.45~GHz FSR microresonator.
	}
	\label{Fig:TransRefl}
\end{figure*}

Supplementary figure \ref{Fig:TransRefl} shows the frequency-dependent transmission and reflection of the photonic chips used throughout this work. \textcolor{black}{We employ frequency comb calibrated laser spectroscopy to perform linear characterization of photonic chips} \cite{del2009frequency}.
\textcolor{black}{We use resonators with the following specs: 2.45 GHz FSR device has 9300 um radius, 800 nm height, 900 nm width waveguide, loaded Q=7.7$\times 10^6$. 9.87 GHz FSR device has 2320 um radius, 900 nm height, 2100 nm width waveguide, loaded Q=10.1$\times 10^6$. 190.7 GHz FSR device has 120 um radius, 950 nm height, 1800 nm width waveguide, loaded Q=4.8$\times 10^6$.}
The FMCW LiDAR experiment is carried out using a microresonator with a free-spectral range (FSR) of 190.7~GHz. The reflection response of the photonic chip is composed of the backreflection of the chip facets and from the microresonator. The former induces sinusoidal modulations of the chip transmission and reflection. The latter leads to normal-mode splitting of the resonances \cite{kippenberg2002modal} a resonant backreflection. A typical Fano shape is observed as the two backreflection pathways interfere, modulating the on resonance response and limiting the injection locking bandwidth for many resonances. However, due to the very high Q of the resonators, even minuscule reflections are sufficient to achieve tight injection locking of the laser \cite{galiev2020injection}. We also note that the wavelength-sized gap and the high refractive index of the InP laser diode will change the phase and amplitude of the sinusoidal modulations compared to the measurement with two lensed fibers.

\textcolor{black}{
\subsection*{Thermal management of the hybrid integrated laser}
The thermal management of the packaged hybrid integrated system is as follows: The laser chip dissipates around 1.5 W and would be mounted on a TEC with active temperature control. The photonic \SiN chip would be mounted on a separate TEC for temperature stabilization. The AlN actuator does dissipate 240 nW at DC and 2.9 $\mu$W at 1 MHz. The DC value is based on our measurement of a 2 nA leakage current at 120V. The AC value is calculated from the equivalent series resistance (ESR) of the AlN dielectric film of a 112 $\mu$m disc actuator with thickness 1 $\mu$m and loss tangent $2.1\times 10^{-3}$. We do not observe a significant red-shift of the microcavity resonance upon starting the AlN actuation.}

\textcolor{black}{
\subsection*{Optical cross correlation measurement of laser phase noise}
At high offset frequencies, the sensitivity of laser frequency noise measurement by heterodyne beat spectroscopy can be limited due to the white phase noise detection floor of photon shot noise that is enhanced by the factor $f^2$ in the phase noise to frequency noise conversion. Optical cross correlation is the technique of choice for the characterization of lasers \cite{xie2017phase} and RF oscillators \cite{rubiola2000correlation} with the highest coherence. The optical setup is depicted in Fig.~3 of the main manuscript. We use free-running Toptica CTL lasers as reference lasers, and the two heterodyne beat notes are recorded using balanced photoreceivers (New Focus 1817). The lasers and photodiodes use independent power supplies to minimize correlated noise sources. The reference lasers are both red-detuned around 20 MHz with respect to the self-injection locked laser. The phase of the heterodyne beat signals are extracted using Hilbert's transform
\begin{equation}
	\Phi_\textrm{meas} (t) = \arg{\mathcal{H} \left\{ U(t) \right\}},
\end{equation}
where $U(t)$ denotes the measured voltage of the heterodyne beat note. The carrier frequency of the beat notes is subtracted by subtracting a linear fit of $\Phi_\textrm{meas} (t)$. The single-sided power spectral densities (PSD) of the two beatnotes and the cross power spectral density (CPSD) are calculated using MATLAB. At higher offset frequencies, the traces are segmented prior to Fourier transform and averaged after Fourier transform, and the results are depicted in Fig.~\ref{Fig:FigSI_CPSD}. The real part of the optical cross correlation signal contains the common-mode noise of the two beatnotes, which is the phase noise of the self-injection locked laser and the suppression of uncorrelated noises such as reference laser noise and photon shot noise reduce as $\sqrt{N}$, where $N$ is the number of averages. We plot both the positive and the negative part of the CPSD to inspect for possible anti-correlated noise sources that would give a false and lower estimate of the phase noise \cite{gruson2020artifacts}. From this, we can conclude that optical cross correlation provides a precise estimation of the SIL laser phase noise for offset frequencies between 10~kHz and 5~MHz. The lower limit is given by insufficient averaging to remove the excess phase noise and spurious anti-correlations of the reference lasers. The upper limit is given by the carrier frequency of the beat notes and the Nyquist sampling limit of the digitizer card.
\begin{figure*}[htbp]
	\centering
	\includegraphics[width=0.75\columnwidth]{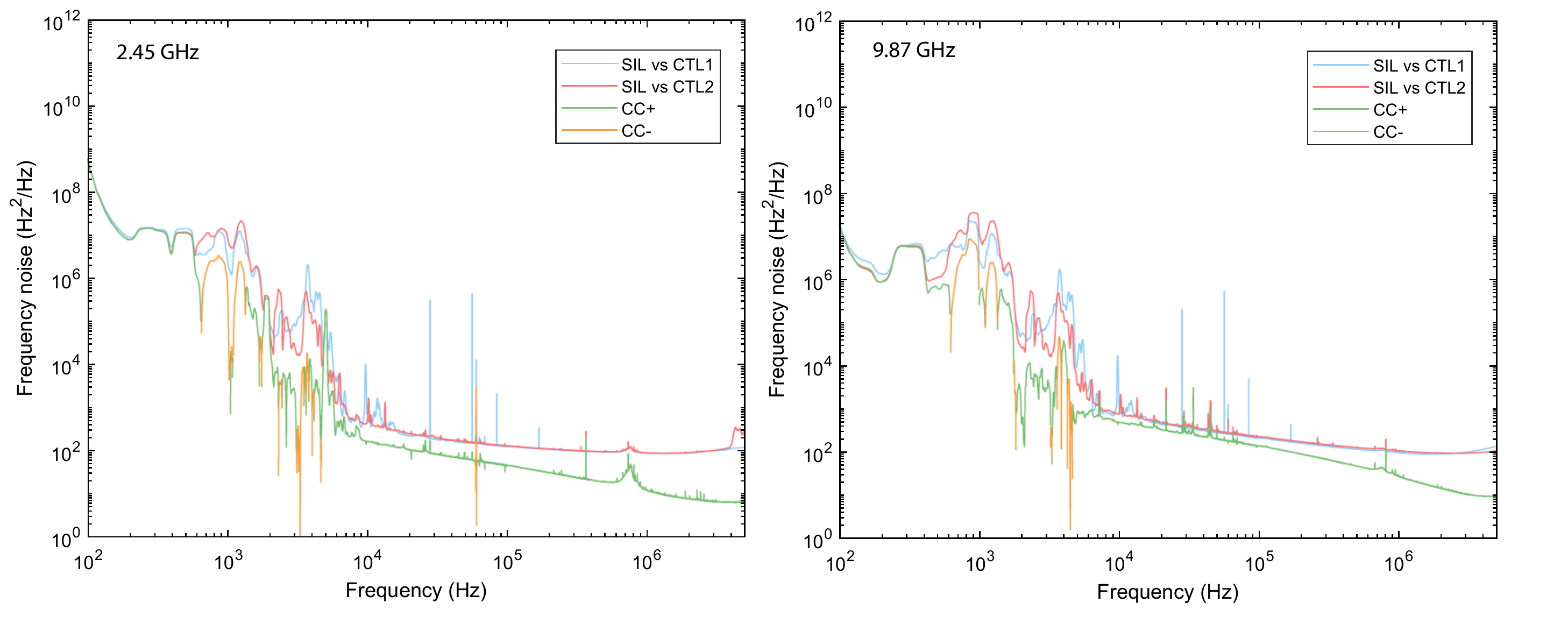}
	\caption{
		\textcolor{black}{\textbf{Optical cross correlation measurement.} Single-sided PSD and CPSD for optical phase noise measurement. The single-sided PSD of the two beatnotes for optical cross-correlation is depicted in blue and red. The positive part of the real part of the CPSD (common-mode noise) is depicted in green, and the negative part (spurious anti-correlation) is depicted in orange.
	}
}
	\label{Fig:FigSI_CPSD}
\end{figure*}
}

\subsection*{Frequency noise measurement results}

\begin{figure*}[htbp]
\centering
l\includegraphics[width=1\columnwidth]{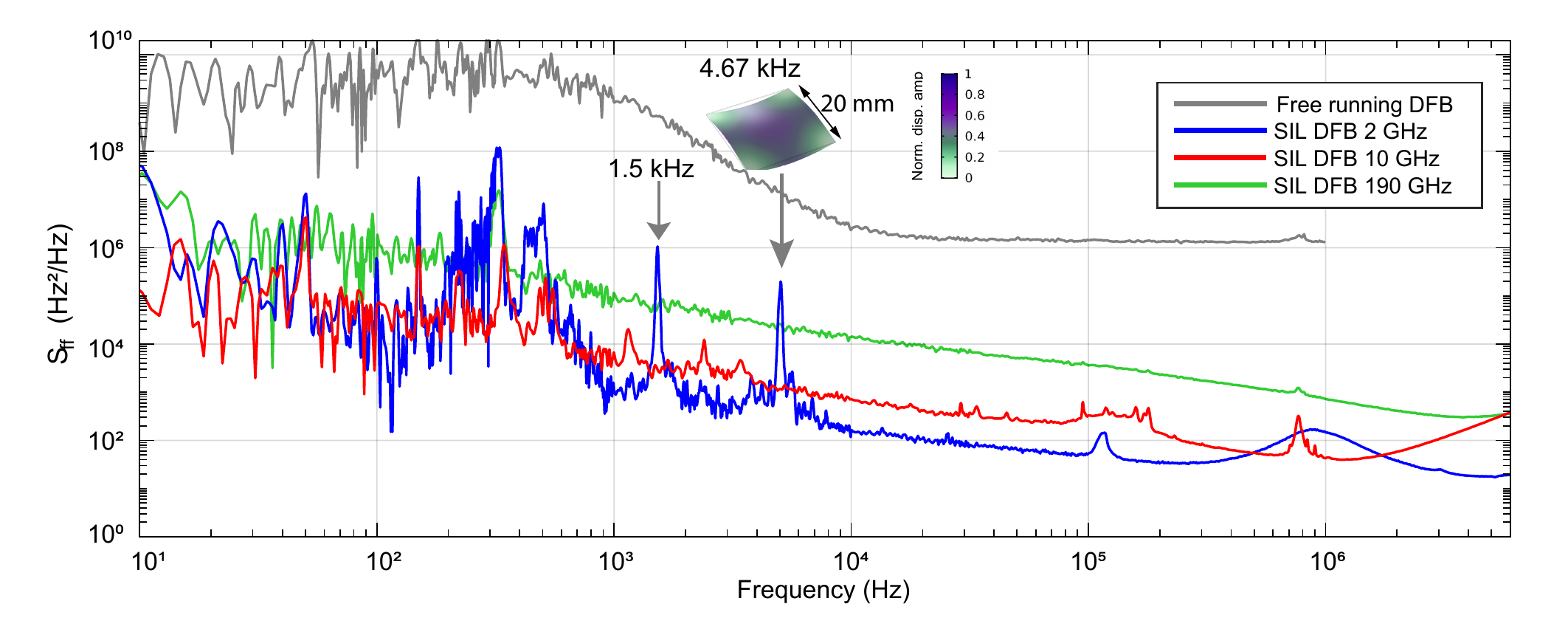}
\caption{
\textbf{Frequency noise of the hybrid integrated laser measured with the single reference laser.} The colors correspond to the colors in the main manuscript Fig.~2. We mainly attribute excess phase noise at 100--700~Hz to vibrations of the experimental setup. The peak at 1.5~kHz is the technical noise from a vacuum pump placed in the lab. We attribute the peak at 5~kHz  to the excitation of the mechanical contour mode of the square 20~mm \SiN chip. The peaks at 100-200 kHz are attributed to the reference laser, and a broad peak around 1~MHz is identified as the servo-bump of the reference laser PDH lock.
}
\label{Fig:SI_FN_full}
\end{figure*}

\noindent The setup and methods for frequency noise measurement are described in the Methods section of the main manuscript. Fig.~\ref{Fig:SI_FN_full} shows the full plot for the frequency noise of the hybrid integrated laser highlighting the low-offset frequency sections. The self-injection-locked laser exhibits several frequency noise spikes in the range of 100--700~Hz, which we mainly attribute to the acoustic noise: vibrations of the experimental setup for the coupling between the laser chip and the Si$_3$N$_4$ photonic chip. We confirmed this by mechanical noise measurements with an accelerometer. With vibration cancellation and packaging, we expect an improvement of the noise level at frequency offsets $<$1~kHz. The difference in sensitivity between the different resonators towards the acoustic noise is related to the different chip size and clamping forces as well as the gap distance between the laser diode and the chip at the optimal phase setting. For the 2.45-GHz-FSR resonator, we observe additional noise peaks at 1.5~kHz, which we attribute to the technical noise from a vacuum pump placed in the lab, and at 5~kHz, which we attribute to the excitation of mechanical contour modes of the Si$_3$N$_4$ chip ($20\times20$ mm size). We confirmed our assumption with FEM simulation. Normalized displacement and frequency are visualized in the inset to Fig.~\ref{Fig:SI_FN_full}. By reducing the Si$_3$N$_4$ chip size and the resonator footprint by using spiral or meander-shaped resonators, we can eliminate most of these noise peaks while simultaneously decreasing the linewidth due to the favorable scaling of TRN noise with the cavity mode volume.

\begin{figure*}[htbp]
\centering
\includegraphics[width=1\columnwidth]{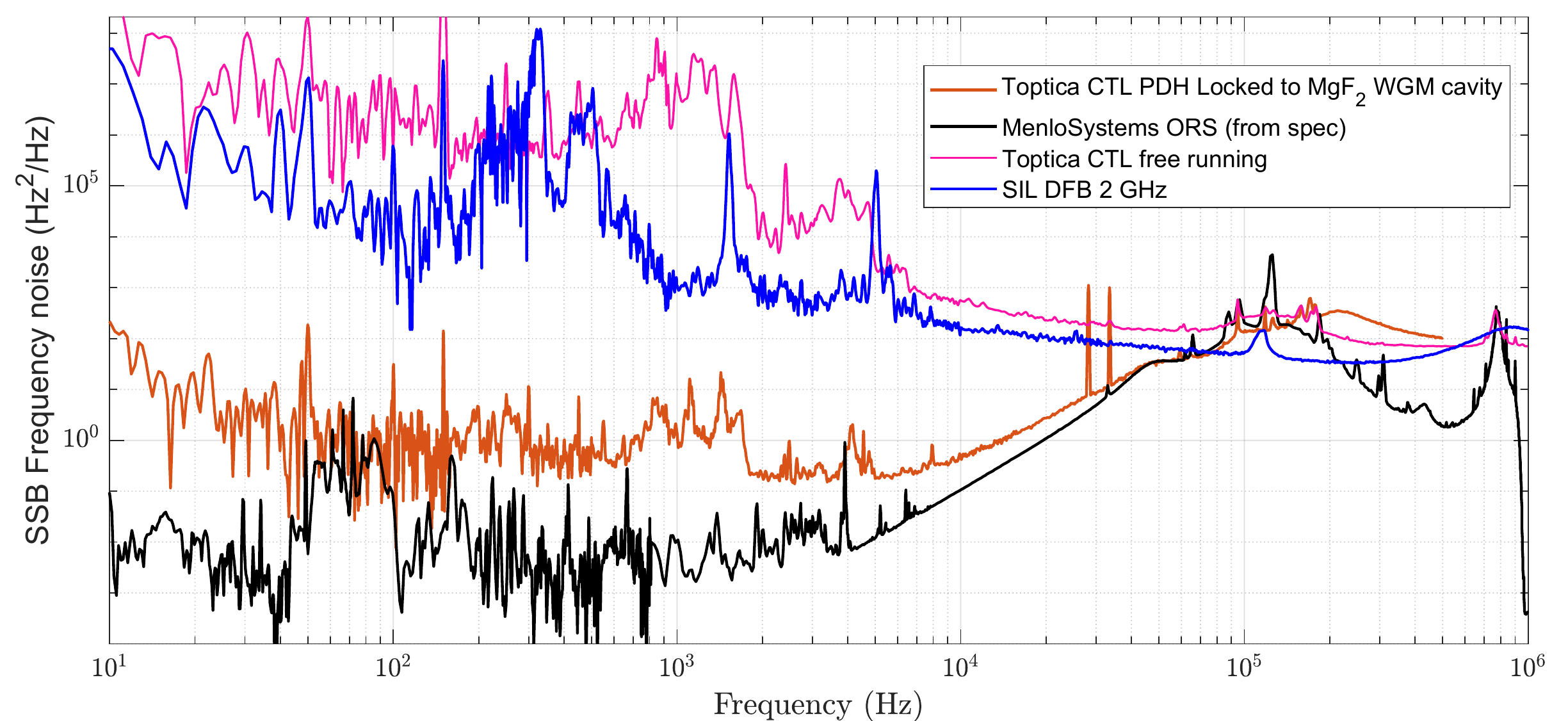}
\caption{
\textcolor{black}{\textbf{Frequency noise of the reference lasers.} (black) MenloSystems ORNS data from specs; (red) Toptica CTL locked to crystalline WGM resonator using PDH technique used as a reference for Fig.~2 of the main Manuscript; (pink) free-running Toptica CTL used as a reference for Fig.~3(d) of the main Manuscript; (blue) SIL DFB 2 GHz FSR.
}
}
\label{Fig:SI_reference_lasers}
\end{figure*}

\textcolor{black}{Fig. \ref{Fig:SI_reference_lasers} presents frequency noise measurement for all reference lasers used in our work. For frequency noise measurements of SIL DFB (Fig.~2 of the main manuscript) we used as a reference laser Toptica CTL PDH-locked to crystalline MgF$_4$ resonator. Frequency noise of PDH locked Toptica CTL is presented in red trace in Fig. \ref{Fig:SI_reference_lasers}, it was measured by heterodyne beat with MenloSystems ORNS (Fig.~\ref{Fig:SI_reference_lasers} black trace). To measure tuning linearity and range for Fig.~3 of the main manuscript we used free-running Toptica CTL (no PDH lock). The frequency noise of Toptica CTL is presented in pink trace in Fig. \ref{Fig:SI_reference_lasers}}.

\textcolor{black}{As shown in Fig.~5 of the SI loaded cavity linewidths are approximately the same across all resonances of the particular \SiN device inside the DFB tuning range 1555-1556.5 nm. Thus, the noise suppression factor due to the laser self-injection locking to different cavity resonance would not vary by more than a factor of 4. Thus, we can choose any microresonator resonance inside DFB current tuning range for laser self-injection locking with proper feedback phase adjustment to obtain a low noise performance at different wavelengths.}

\subsection*{Intensity noise measurement}
\noindent The relative intensity noise (RIN) is important to characterize the laser performance for numerous applications, especially in optical communication systems. The RIN is measured with a DC-coupled photodiode for the case of the 9.87~GHz microresonator. A large DC block capacitor (100~$\mu$F) with frequency cut-off below 1~Hz is inserted to separately measure the DC photocurrent. The noise fluctuation is measured with the same electrical spectrum analyser that was used for the frequency noise measurement. Compared with the RIN of the laser in free-running condition, the injection-locked laser exhibits a noticeable RIN increase at frequencies below 10~kHz, which is caused by the vibrations of mechanical stages that constitute the coupling setup between the laser chip and the Si$_3$N$_4$ photonic chip. With better vibration cancellation or compact and rigid component packaging, the excess frequency noise and RIN deterioration at low offset frequencies can be avoided.

No relaxation oscillations peaks were observed in our RIN measurements for the free-running and self-injection locked DFB. According to the simulations presented above, a relaxation oscillation frequency of the DFB laser should be above 10~GHz both in the free-running and in the SIL state.

\begin{figure*}[htbp]
\centering
\includegraphics[width=0.75\columnwidth]{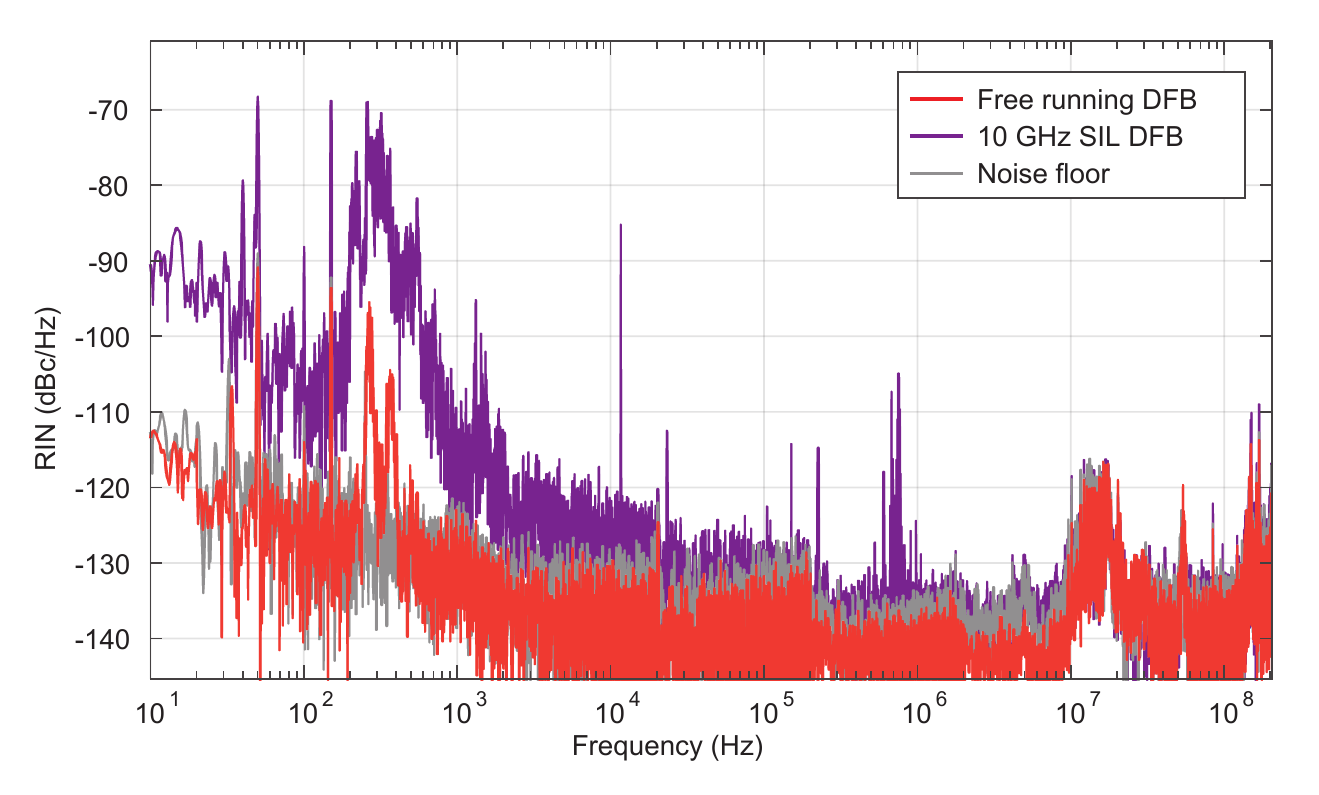}
\caption{
\textbf{Laser RIN measurement.} DFB laser is free-running (red) or self-injection locked to the 9.78 GHz \SiN~microresonator (violet). Noise at 100--700~Hz we mainly attribute to vibrations of the experimental setup.
}
\label{Fig:FigSI_RIN}
\end{figure*}

\textcolor{black}{
\subsection*{Phase noise measurement of frequency-modulated laser}
\noindent The phase of a rapidly frequency-modulated laser, such as a FMCW laser source, is defined as the time integral of over the target frequency-modulation.
\begin{equation}
\Phi (t) = \Phi (0) + \omega_0 \cdot t + \int_{0}^{t} \Delta\omega(t') dt'
\end{equation}
For a direct comparison of the laser phase noise in the CW and FMCW operation modes, we can extract the modulated laser phase from the heterodyne beat spectroscopy experiment depicted in Fig. 3 of the main manuscript using Hilbert's transform. A Toptica CTL laser was used as a reference in the experiment. Residual intensity noise of the lasers is subtracted by a 10~MHz low pass filtering of the beat note prior to phase extraction. This does not influence the phase noise extraction because the lowest carrier frequencies of the beat note are around 500~MHz. The top row of Fig. \ref{Fig:FigSI_RampPhaseNoise} shows the phase deviation $\Delta\Phi$ between the measured phase and the phase of an ideal triangular frequency modulation, which is extracted from our fit of the frequency-modulation after subtraction of a linear polynomial to eliminate the constant phase and frequency offsets. The phase noise power spectral densities are depicted in the lower row of Fig. \ref{Fig:FigSI_RampPhaseNoise}. In each case, the measurement duration corresponds to between seven and ten periods of the triangular chirp.
\begin{figure*}[htbp]
	\centering
	\includegraphics[width=\columnwidth]{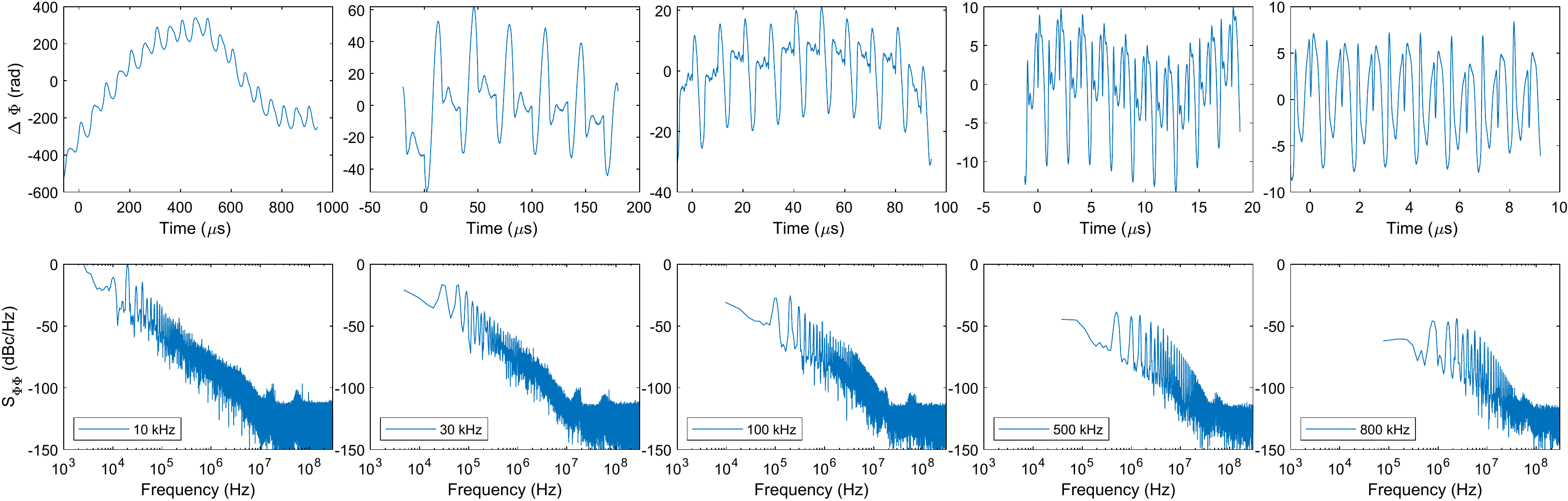}
	\caption{
		\textcolor{black}{\textbf{Phase noise of triangular frequency-modulated laser}. The top row depicts the phase difference between the measured heterodyne beat note and the ideal phase curve of a triangular frequency-modulated laser. The bottom row depicts the single-sided power spectral density of the phase deviation. The legend indicates the repetition frequency of the triangular ramp chirps.
	}
}
	\label{Fig:FigSI_RampPhaseNoise}
\end{figure*}
At low offset frequencies, the phase noise PSD is dominated by the residual nonlinearities of the odd and even harmonics of the repetition frequency of the triangular chirp. These could be further reduced by pre-distortion \cite{feneyrou2017frequency} of the driving signal or by phase locking to an imbalanced Mach-Zehnder Interferometer \cite{roos2009ultrabroadband}. For chirp repetition frequencies between 10~kHz and 30~kHz, we find phase noise of around -80~dBc/Hz around 1~MHz offset, which corresponds to $10^{4}$~Hz$^{2}$/Hz and is within a factor of ten from the CW case for self-injection locking to the 190.7~GHz microresonator. We attribute the increase of the phase noise to a decrease of the phase noise reduction close to the edge of the locking bandwidth.
}

\subsection*{Free running DFB characterization}
\begin{figure*}[htbp]
\centering
\includegraphics[width=0.6\columnwidth]{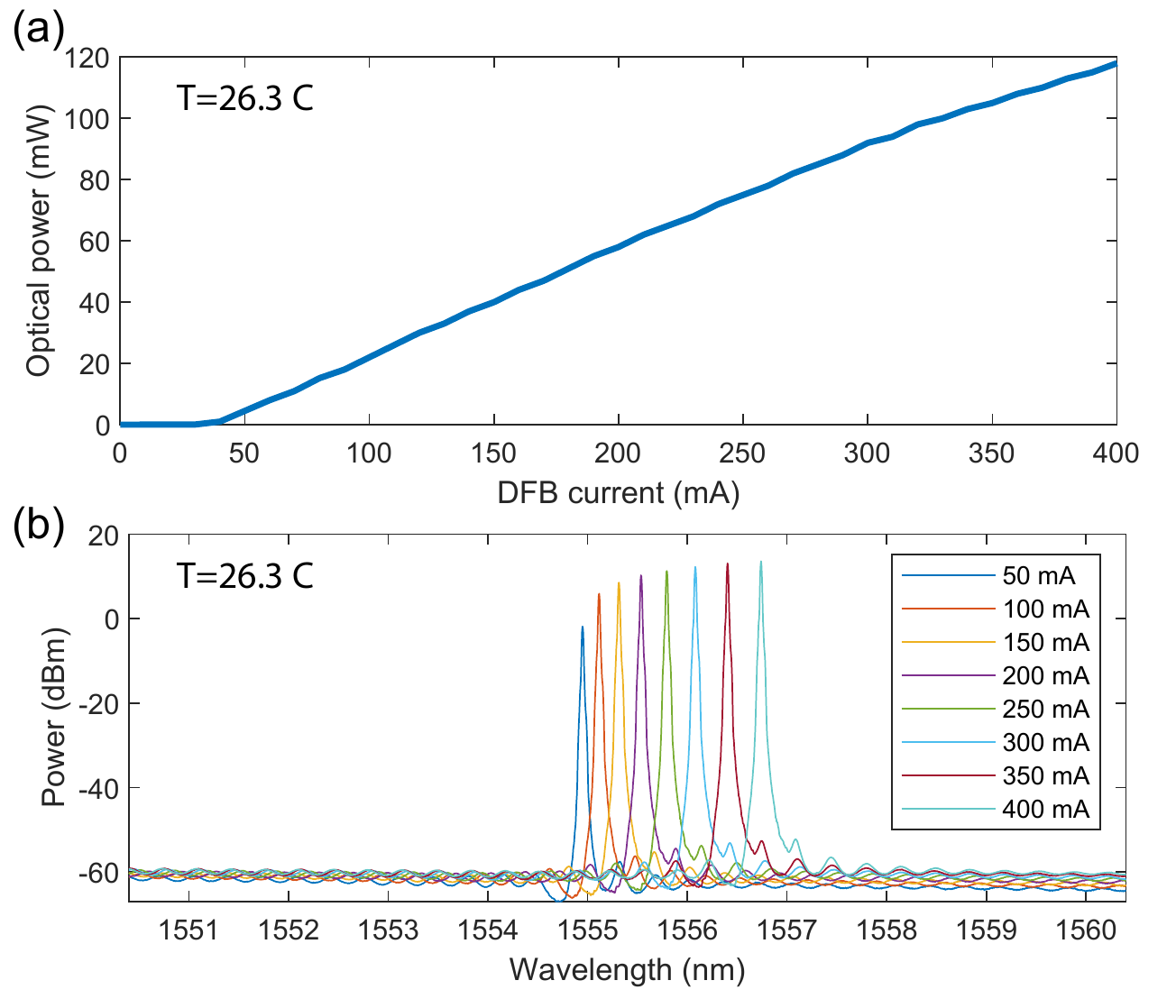}
\caption{
\textcolor{black}{\textbf{Free running DFB characterization.} a) Laser diode free-space optical power vs. diode current. b) Free-running DFB optical spectra at different driving currents.}
}
\label{Fig:SI_DFB_characterization}
\end{figure*}

\textcolor{black}{
We perform characterization of the free running DFB laser (no \SiN ~chip). Figure \ref{Fig:SI_DFB_characterization} (a) shows a free-space optical power vs. DFB current. Figure \ref{Fig:SI_DFB_characterization} (b) presents optical spectra of DFB at different driving currents.}

\textcolor{black}{We perform DFB current tuning linearity measurement using a similar experimental setup and analysis method as in Figure 3 of the main manuscript. We drive the laser diode with 155 mA of current with a 3.5 mA triangular chirp at 1 kHz and 10 kHz rates. Higher current modulation frequencies are limited by the current controller bandwidth. We choose 3.5 mA current ramp amplitude to get 1.2 GHz optical frequency excursion similar to the SIL tuning case. Optical frequency excursion is 1.2 GHz Figure \ref{Fig:SI_DFB_characterization} shows the time-frequency spectrogram of the heterodyne beat-notes with fixed frequency reference laser (Toptica CTL) for 1 kHz and 10 kHz chirping rates. The bottom row reveals RMS nonlinearities 30 MHz for 1 kHz tuning rate and 120 MHz for 10 kHz. Such nonlinearities are two orders of magnitude higher than the values measured in self-injection locked laser regime with piezoactuator voltage tuning. Thus, we demonstrate that the self-injection locking to the photonic microresonator with tuning provided the integrated piezoactuator can significantly improve the laser tuning linearity and speed.}

\begin{figure*}[htbp]
\centering
\includegraphics[width=0.6\columnwidth]{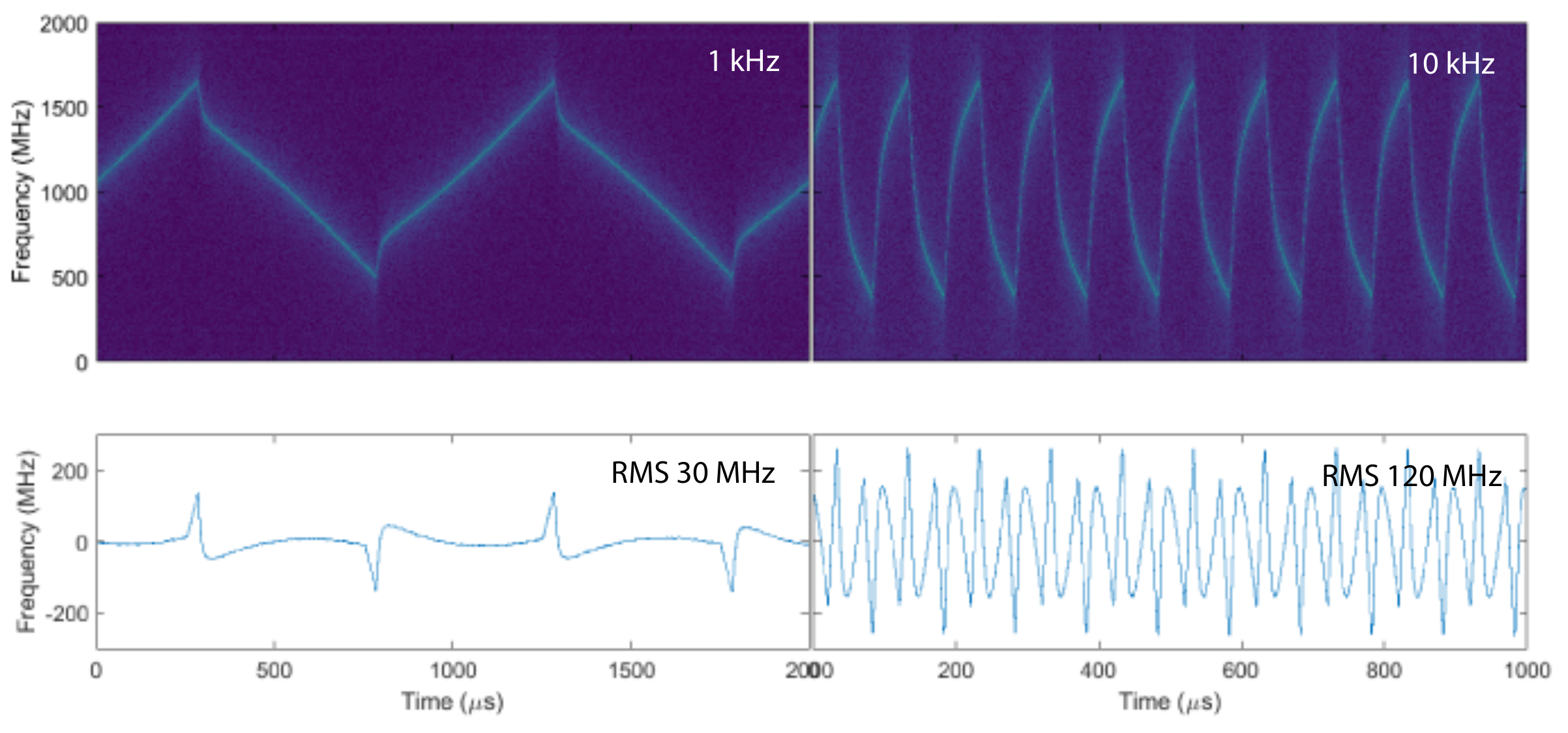}
\caption{
\textcolor{black}{\textbf{Free running DFB current tuning linearity.} Time-frequency spectrogram of the heterodyne beatnotes for 1 kHz and 10 kHz triangular chirp frequencies. Bottom row: Residual of least-squares fitting of the time-frequency traces with symmetric triangular chirp pattern.}
}
\label{SI_DFB_current_tuning}
\end{figure*}

\subsection*{Frequency noise and tuning comparison between different compact lasers}
\begin{figure*}[htbp]
\centering
\includegraphics[width=\columnwidth]{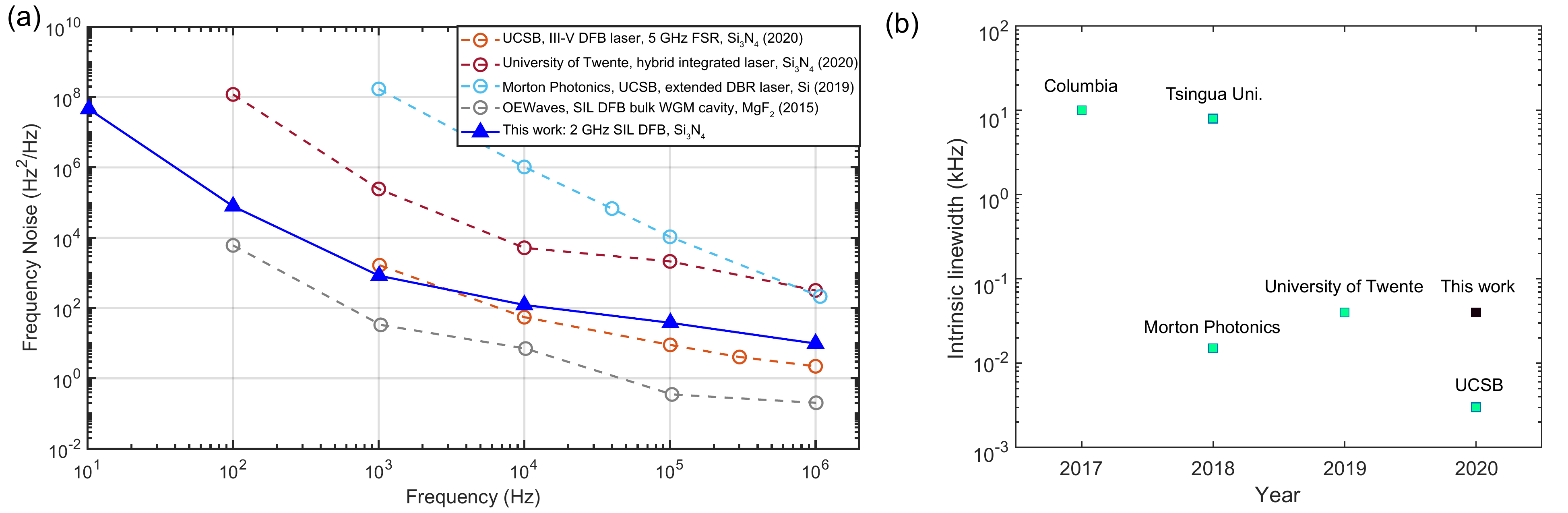}
\caption{
\textbf{Frequency noise of different integrated lasers.} a) Frequency noise of several contemporary hybrid and heterodyne integrated lasers. Citations are found in the SI text. b) Intrinsic linewidth of different lasers reported by universities or companies.
}
\label{Fig:SI_FN_compare}
\end{figure*}

Fig. \ref{Fig:SI_FN_compare} compares frequency noise of hybrid and heterogeneous integrated lasers reported in recent literature. The material systems used include glass (Morton Photonics \cite{morton2018high}), strong-confinement Si$_3$N$_4$ (Columbia \cite{stern2017compact}), weak-confinement Si$_3$N$_4$ (UCSB \cite{jin2020hertz} \& \cite{xiang2019ultra}, \textcolor{black}{University of Twente} \cite{boller2020hybrid}). The list includes both heterogeneous- (UCSB \cite{xiang2019ultra}) and hybrid-integrated laser systems. Lowest reported value of frequency noise in similar systems was achieved in packaged system with a laser self-injection locked to a bulk crystalline MgF$_2$. \cite{liang2015ultralow}

\textcolor{black}{Table \ref{table_laser_comparison} presents a comparison of different tunable laser systems in terms of the frequency tuning range, tuning rate, linearity, optical output power, frequency white noise floor.}

%\begin{sidewaystable} [htbp]
\begin{table} [htbp]
	\centering
		%\captionsetup{format=tablenocaption}
		\begin{tabular}{||c|c|c|c|c|c|c|c||}
			
			\hline
			\multicolumn{1}{|p{2cm}|}{\centering Author} & \multicolumn{1}{|p{2cm}|}{\centering System} & \multicolumn{1}{|p{2cm}|}{\centering Frequency white noise floor} & \multicolumn{1}{|p{2cm}|}{\centering Tuning speed, $f_{mod}*B$} &
			\multicolumn{1}{|p{2cm}|}{\centering Tuning rate ($f_{mod}$)} &
			\multicolumn{1}{|p{2cm}|}{\centering Tuning range (B)} &
			\multicolumn{1}{|p{2cm}|}{\centering Linearity} &
			\multicolumn{1}{|p{2cm}|}{\centering Optical output power (mW)}\\
			\hline
			Yoo, S et al.\cite{Yoo:17} & DS-DBR & 5 MHz & 1.54E+16 & 1.23 MHz & 12.5 GHz & \multicolumn{1}{|p{2cm}|}{\centering Low, Wavelength  switching} & 6 \\
			\hline
			Nunoya, N. et al.\cite{Nunoya5749683} & TDA-DFB & n/a & 3.75E+15 & 5 kHz & 0.75 THz & \multicolumn{1}{|p{2cm}|}{\centering Low, Wavelength  switching} & 20 \\
			\hline
			Y. Fan et al.\cite{fan2020hybrid} & \multicolumn{1}{|p{2cm}|}{\centering \SiN Vernier filter} & 40 Hz & n/a & \multicolumn{1}{|p{2cm}|}{\centering Integrated heaters} & 8.7 THz & n/a & 23 \\
			\hline
			M. A. Tran et al.\cite{Tran8805353} & \multicolumn{1}{|p{2cm}|}{\centering Heterogeneous Si/InP} & 220 Hz & n/a & \multicolumn{1}{|p{2cm}|}{\centering Integrated heaters} & \multicolumn{1}{|p{2cm}|}{\centering 30 GHz (continuous)} & Low & 3.5 \\
			\hline
			C.V. Poulton et al.\cite{Poulton:17} & DFB & n/a & 6.4E+12 & 100 Hz & 64 GHz & High & 1 \\
			\hline
			N. Satyan et al.\cite{Satyan:09} & DFB SCL & 1 MHz & 1E+14 & 1 kHz & 100 GHz & High (OPLL) & 40 \\
			\hline
			Zhang et al.\cite{Zhang:19} & VCSEL & n/a & 7.75E+14 & 5 kHz & 155 GHz & \multicolumn{1}{|p{2cm}|}{\centering High (ILC pre - distortion)} & n/a \\
			\hline
			Measured EPFL & \multicolumn{1}{|p{2cm}|}{\centering Toptica CTL} & 90 Hz & 2.5E+14 & 100 Hz & 2.5 THz & Low & 30 \\
			\hline
			Measured EPFL & \multicolumn{1}{|p{2cm}|}{\centering Koheras Adjustik E15}& <7 Hz & 2E+13 & 20 kHz & \multicolumn{1}{|p{2cm}|}{\centering 1 GHz (piezo)} & Low & 1.5 \\
			\hline
			Spec sheet & \multicolumn{1}{|p{2cm}|}{\centering Lumentum NPRO 126} & <5 kHz & 9E+11 & 30 kHz & 30 MHz & n/a & >100\\
\hline
			Okano et al.\cite{Okano:20} & VCSEL & <1 MHz & 1.1E+17 & 10 kHz & 11 THz & \multicolumn{1}{|p{2cm}|}{\centering High (k-point sampling)} & 1\\
			\hline
			DiLazaro et al.\cite{DiLazaro:18}& \multicolumn{1}{|p{2cm}|}{\centering 12 stitched DFBs} & 3 MHz & 1.8E+15 & 330 Hz & 5.56 THz & \multicolumn{1}{|p{2cm}|}{\centering High (with linearization)} & 12.7\\
\hline
			This work & SIL DFB & 25 Hz & 1.6E+15 & 800 kHz & 2 GHz & \multicolumn{1}{|p{2cm}|}{\centering High} & 1.5\\
			\hline
			
		\end{tabular}
\caption{Performance comparison of laser systems.}
\label{table_laser_comparison}
	
\end{table}

\subsection*{LiDAR data analysis}

The collected data were processed in order to obtain the location of the objects comprising the studied scene. First, the padding of oscillogram was carried out to align the zero point due to different time delays between the oscilloscope trigger signal and LiDAR signal path. The second step was to perform short-time Fourier transform (STFT) of the oscillograms from the target and the reference MZI. The Blackman-Harris window function was chosen for STFT with the window size set to half of the laser frequency chirping period. Next, time-frequency spectrograms were analysed, and the maximal values of each timeslice were identified for the subsequent conversion of the corresponding frequency points into the distance units. An alternative approach consisted of the application of a Gaussian fitting procedure for searching the maximal spectrum values. After that, the noise filtering stage followed that cut off the identified peaks lower than a given value of 10 dB that is attributed to noise or low reflection. Next, the galvo-mirror angular coordinates linearization was applied to reduce the digitalization noise smoothing the beam scanning pattern to the linear ramp shape. Finally, a three-dimensional spatial distribution of the detected points was plotted with distance-based coloring. To achieve a greater quality of image, additional filtering or N-points-averaging can be performed for a point cloud to reduce a discretization error which is due to 1) a LiDAR resolution of \textcolor{black}{12.5 cm} and 2) the distance ambiguity related to the laser frequency Doppler shift due to a fast rotation of galvo mirrors (60 Hz in our case). No additional filtering has been applied to an image presented in Fig. 4 of the main manuscript. We attach interactive Python code (in Zenodo data repository) for LiDAR data processing with a detailed step-by-step descriptions.

\subsection*{Soliton microcomb generation with ultralow noise repetition rate signal}

%%%%%%%%%%%%%%%%%%%%%%%%%%%%%%%%%%%%%%%%%%%%%%%%%%%%%%%%%%%%%%%%%%%%%%%%%%%%%%%%

\begin{figure*}[htbp]
	\centering
	\includegraphics[width=1\columnwidth]{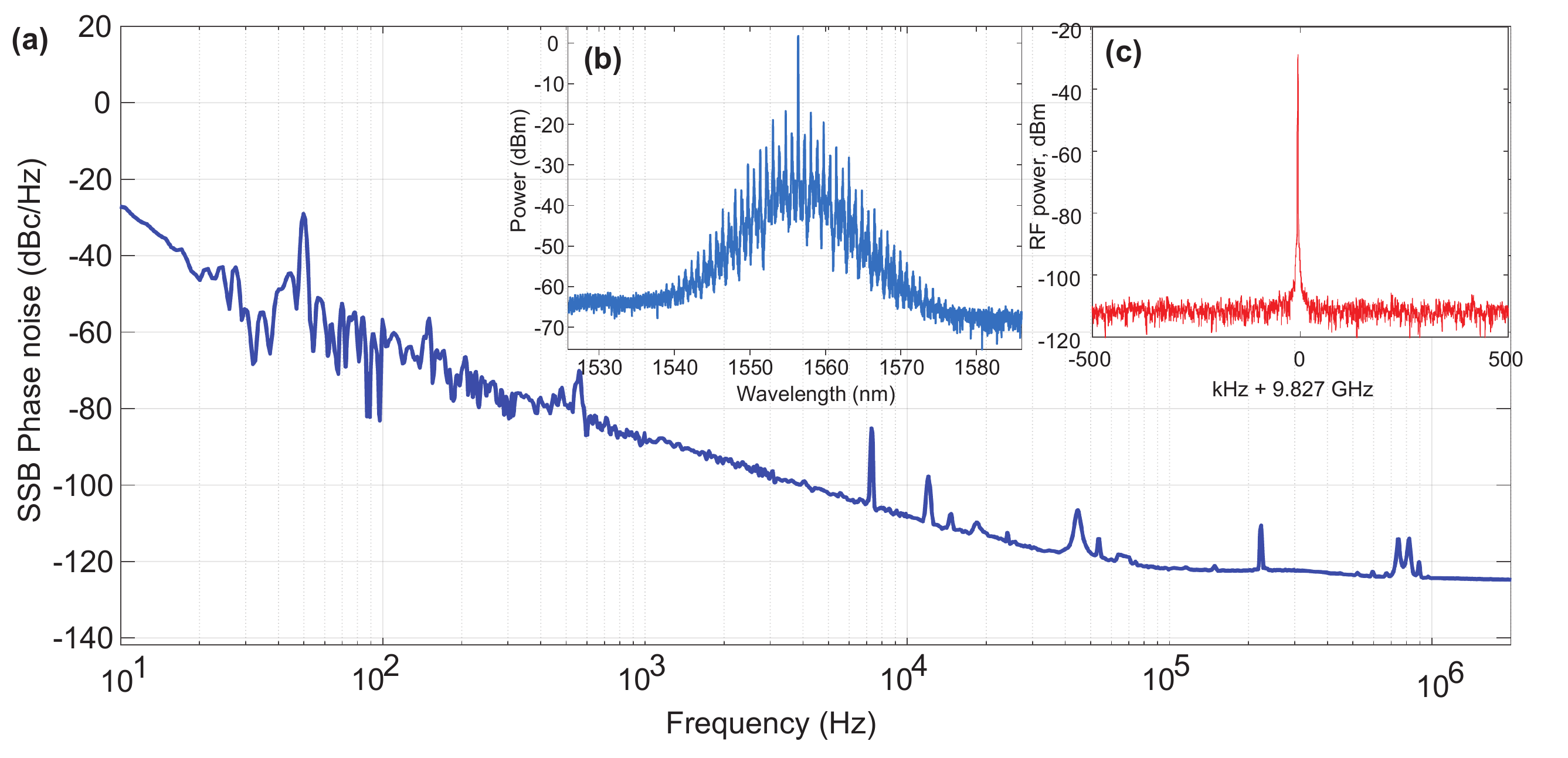}
	\caption{
		\textbf{Microcomb generation with the DFB laser injection-locked to the pumped mode}. (a) Single-sideband phase noise spectrum of the repetition rate of soliton microcomb generated with self-injection locked DFB. (b) A multi-soliton microcomb optical spectrum generated with the hybrid integrated laser system. (c) Soliton repetition rate signal at 9.8 GHz.
	}
	\label{Fig:FigSI_comb}
\end{figure*}

Owing to the ultra-low optical loss resulted from the photonic Damascene reflow process and the engineered anomalous dispersion ($D_2/2\pi = 12.8$~kHz) that supports dissipative Kerr soliton (DKS) generation in the telecom band, the InP laser power is sufficient to generate stable DKS microcomb in the Si$_3$N$_4$ photonic chip. To produce DKS microcomb directly from the hybrid integrated laser system, a Si$_3$N$_4$ microresonator with an FSR of $\sim10$~GHz is used for this experiment. Soliton steps are observed in the self-injection-locking state as we sweep the InP laser frequency over a mode resonance of the Si$_3$N$_4$ microresonator.
We can operate the system in the DKS regime by stopping the current sweep in the spectral region of soliton existence.
Fig. \ref{Fig:FigSI_comb} (b) displays the microcomb spectrum, revealing that the microcomb is in a multi-soliton state. We note that due to the indeterministic nature of soliton excitation \cite{herr2014temporal} and the relatively large size of the microresonator in this experiment, the single-soliton state is rarely obtained unless extra soliton number switching \cite{guo2017universal} is implemented. We also measure the phase noise of the DKS repetition rate by detecting the soliton train with a fast photodetector and then using a phase noise analyser \textcolor{black}{(Rohde \& Schwarz FSW43)} to measure the single-sideband phase noise of the repetition rate signal. Shown in Fig.\,\ref{Fig:FigSI_comb} (a), the demonstrated phase noise level is lower than those obtained with a fibre laser (Koheras), and an external-cavity diode laser (ECDL) with a linewidth of $\sim10$ kHz in our previous work \cite{liu2020photonic}. The repetition rate phase noise level of the microcomb generated with the hybrid integrated laser exhibit a noise reduction of up to 20 dB in the frequency range of 10~Hz -- 10~kHz, reaching $-25$ and $-60$ dBc/Hz at the offset frequencies of 10 and 100 Hz respectively. Owing to the excellent noise performance of the self-injection-locked laser, to the best of our knowledge, the phase noise of the soliton repetition rate shows the best value with DKS microcombs based Si$_3$N$_4$ microresonators, whose noise level is often limited by the pump laser's RIN and phase noise \cite{liu2020photonic}. This low phase noise that is attributed to the combination of the self-injection locking technique and the microcomb generation technique not only relieves the microcomb system from using bulk and expensive laser sources but also may enable wide applications of microcombs in microwave photonics due to its superior performance.

\subsection*{Mode-cancellation schemes for actuation bandwidth extension}

Contour modes of an integrated MEMS-photonic chip have their fundamental frequency defined by its in-plane dimensions, and the operational modes can be broadly categorized into flexural and bulk modes \cite{piazza2006piezoelectric}. At first, we develop a difference-actuation scheme. In this scheme, besides the AlN actuator above the microresonator, an additional AlN actuator with the same geometry is fabricated adjacent to the microresonator but with no Si$_3$N$_4$ microring resonator below it. During actuation, the two actuators are driven by the same frequency with an identical amplitude but different phases to cancel the actuation of the photonic chip mechanical modes. As a result, while the stress-optical effect exerted on the microresonator is the same, the detrimental flexural modes arising due to transverse standing waves get effectively suppressed. Fig. 3 (e) in the main manuscript shows the measured response of the conventional single-actuator configuration and the difference-actuation configuration. The comparison clearly shows that the latter effectively reduces the amplitudes of the mechanical modes below 1~MHz (green trace), mainly cancelling the flexural modes due to far-field destructive interference. It is clearly observed that the bulk mechanical mode at 967 kHz does not get affected by this scheme. We use the finite-element method to compute the contour modes' profiles of an actual photonic chip with the size of 4.96mm x 4.96mm. In insets of Fig. 3 (e) we show three of these eigenmodes. The eigenmodes at 225 kHz and 490 kHz are flexural modes, whereas the one at 967 kHz is a bulk mode. The actuation voltage derived from a vectorial network analyser (VNA) is applied on the two actuators in an anti-phase fashion, and a laser is frequency-tuned to sit on the side of resonance. The frequency modulation due to the actuation is converted to the intensity modulation of the transmitted laser light that is received by a fast photodetector.

\begin{figure*}[htbp]
\centering
\includegraphics[width=0.4\columnwidth]{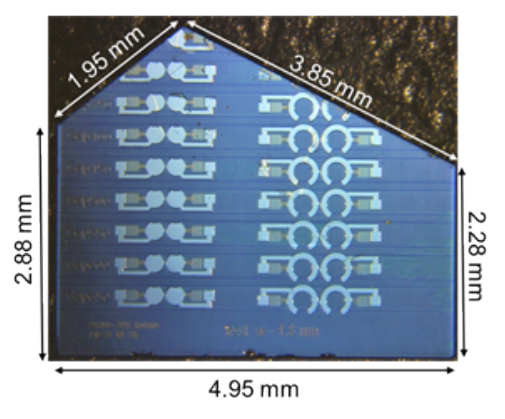}
\caption{\textcolor{black}{\textbf{Photo of the apodized chip.}}}
\label{Fig:FigSI_apodized_chip}
\end{figure*}

\begin{figure*}[htbp]
\centering
\includegraphics[width=1\columnwidth]{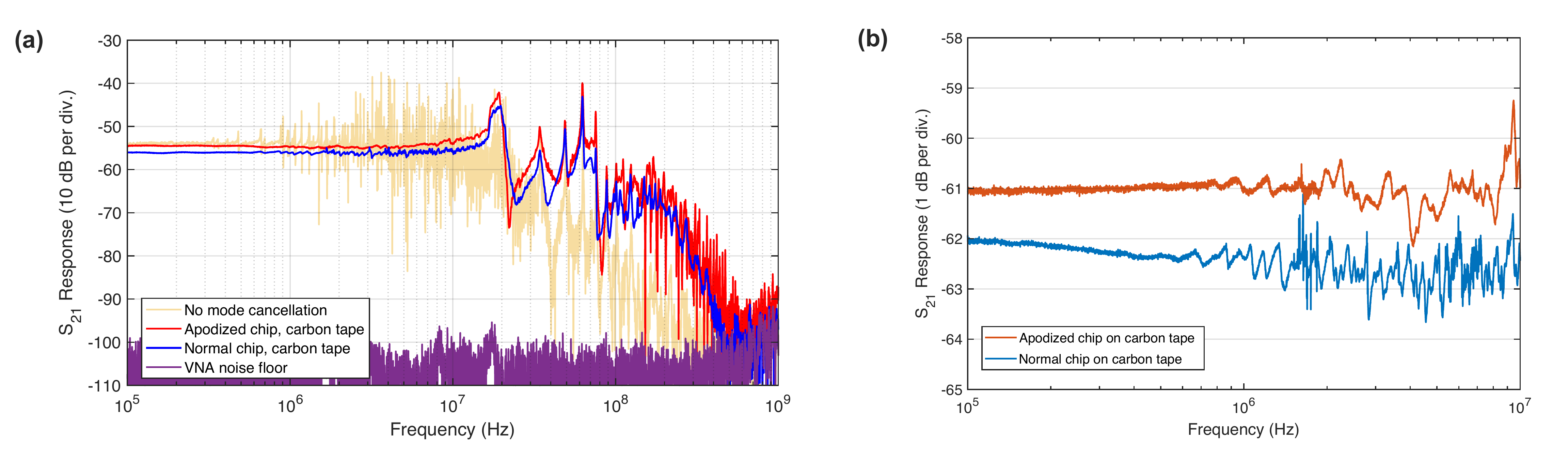}
\caption{
\textbf{S$_{21}$ response showing mechanical resonances of the photonic chip}. (a)  Full spectrum showing suppression of mechanical modes using apodized chip and normal square chip placed over a piece of carbon tape. (b) Apodized chip shows response fluctuation within 1 dB due to effective suppression of bulk mechanical modes as compared to a normal chip.}
\label{Fig:FigSI_ModeCancellation}
\end{figure*}

Next, we implement apodized shape engineering on the photonic chips. Many mechanical modes of relatively low resonance frequencies are flexural modes whose vibrations are caused by transverse standing waves \cite{wu2020mems}. The bulk mechanical modes, whose vibrations are caused by longitudinal standing waves, can be eliminated by judiciously shaping the geometry of the photonic chips \cite{ruby2009piezoelectric, burak2017acoustic}. \textcolor{black}{The apodization was performed by dicing the released chip, see Fig. {\ref{Fig:FigSI_apodized_chip}}}. We observed a reduction in the number of bulk mechanical modes in an apodized photonic chip. We repeat the actuation response measurement, and in the result that is presented in Fig. 3 (e) of the main manuscript, it is shown that the mechanical resonances below  $\sim1.69$ MHz are significantly suppressed (red trace). \textcolor{black}{We confirm first mechanical mode of the apodized chip at 1.69 MHz with FEM simulations}. We further flatten the actuation response by attaching the apodized chip on a piece of carbon tape and then differentially driving the actuators, as explained before. In this way, both the flexural and the bulk mechanical modes are damped up to the first HBAR mode at 17 MHz as shown in Fig. \ref{Fig:FigSI_ModeCancellation} (a). It is also observed that even without using the difference-actuation scheme, all the mechanical resonances of the apodized chip are significantly suppressed till the first HBAR mode, but the difference-actuation scheme on an apodized chip placed over a carbon tape limits the fluctuation of the actuation response within 1 dB, giving the best flattened result as depicted in Fig. \ref{Fig:FigSI_ModeCancellation} (b). This actuation response can significantly improve the linear chirping performance of the system as a LiDAR engine.

\bibliography{citations_SI}